\documentclass[aps,prd,preprintnumbers,tightenlines,floats,twocolumn,nofootinbib,amsmath,amssymb,showpacs,superscriptaddress]{revtex4}
\usepackage{amsmath,amssymb,amsfonts}
\usepackage{graphicx}
\usepackage{enumerate} 
\usepackage{colordvi} 
\usepackage{color} 
\usepackage{bm}

\newcommand{\bfv}{\mbox{\boldmath$v$}}

\newcommand{\bfx}{\mbox{\boldmath$x$}}
\newcommand{\bfk}{\mbox{\boldmath$k$}}
\newcommand{\bfp}{\mbox{\boldmath$p$}}

\newcommand{\bfr}{\mbox{\boldmath$r$}}

\newcommand{\Pdd}{P_{\delta\delta}}
\newcommand{\Pdv}{P_{\delta \theta}}
\newcommand{\Pvv}{P_{\theta\theta}}
\newcommand{\DFoG}{D_{\rm FoG}}

\newcommand{\sigmav}{\sigma_{\rm v}}
\newcommand{\rhom}{\rho_{\rm m}}

\begin{document}
\title{Beyond consistency test of gravity with
redshift-space distortions at quasi-linear scales}

\author{Atsushi Taruya}
\affiliation{Yukawa Institute for Theoretical Physics, Kyoto University, Kyoto 606-8502, Japan}
\affiliation{Research Center for the Early Universe, Graduate School of Science,
The University of Tokyo, Bunkyo-ku, Tokyo 113-0033, Japan}
\affiliation{
Kavli Institute for the Physics and Mathematics of the Universe, Todai Institutes for Advanced Study, the University of Tokyo, Kashiwa, Chiba 277-8583, Japan (Kavli IPMU, WPI)}
\author{Kazuya Koyama}
\affiliation{Institute of Cosmology \& Gravitation, University of Portsmouth, Dennis Sciama Building, Portsmouth, PO1 3FX, United Kingdom}
\author{Takashi Hiramatsu}
\affiliation{Yukawa Institute for Theoretical Physics, Kyoto University, Kyoto 606-8502, Japan}
\author{Akira Oka}
\affiliation{Department of Physics,
The University of Tokyo, Bunkyo-ku, Tokyo 113-0033, Japan}

\date{\today}
\begin{abstract}
Redshift-space distortions (RSD) offer an attractive method to
measure the growth of cosmic structure on large scales,
and combining with the measurement of the cosmic expansion history,
it can be used as cosmological tests of gravity. With the advent of future
galaxy redshift surveys aiming at
precisely measuring the RSD, an accurate modeling of RSD going beyond
linear theory is a critical issue in order to detect or disprove small
deviations from general relativity (GR). While several improved
models of RSD have been recently proposed based on the perturbation theory
(PT), the framework of these models heavily relies on GR. 
Here, we put forward a new PT prescription for RSD in general
modified gravity models. As a specific application, we present theoretical
predictions of the redshift-space power spectra in $f(R)$ gravity model,
and compare them with $N$-body simulations.
Using the PT template that takes into account the effects of both
modifications of gravity and RSD properly, we successfully 
recover the fiducial model parameter 
in $N$-body simulations in an unbiased way. 
On the other hand, we found it difficult to
detect the scale dependence of the growth rate in a model-independent 
way based on GR templates.  
\end{abstract}

\preprint{YITP-13-93}
\maketitle

\section{Introduction}
\label{sec:intro}
Redshift-space distortions (RSD) of galaxy clustering, which appear as
systematic effects in determining the redshift of each galaxy
via spectroscopic measurements and manifestly break statistical isotropy
\cite{Hamilton:1997zq,Peebles:1980},
are now recognized as a sensitive probe of the growth of structure.
Combining the distance measurement of galaxies using the
baryon acoustic oscillations as standard ruler (e.g., \cite{Seo:2003pu,
Blake:2003rh,Matsubara:2004fr,Glazebrook:2005mb}), RSD offers a unique
opportunity to test the theory of gravity on cosmological scales
(e.g., \cite{Linder:2007nu,Guzzo:2008ac,Yamamoto:2008gr,
Song:2008qt,Song:2010bk, Guzik:2009cm, Song:2010fg, Asaba:2013xql}) and help us obtain a deeper understanding of the
current accelerating expansion of the Universe.
This is indeed one of the main goal of on-going and up-coming galaxy
surveys such as the
Baryon Oscillation Spectroscopic Survey of Sloan Digital Sky
Survey III\footnote{{\tt http://www.sdss3.org}},
the WiggleZ survey\footnote{{\tt wigglez.swin.edu.au}},
the Subaru Measurement of Imaging and
Redshifts\footnote{{\tt http://sumire.ipmu.jp/en/}},
the Dark Energy Survey\footnote{{\tt www.darkenergysurvey.org}},
the BigBOSS project\footnote{{\tt bigboss.lbl.gov/index.html}}, and
the ESA/Euclid survey\footnote{{\tt www.euclid-ec.org}},
which will provide precision measurements of the power spectrum
(or correlation function). The late-time cosmic acceleration,
first discovered by the observations of distant type Ia supernovae
\cite{Perlmutter:1998np,Riess:1998cb},
may be the result of a dark energy which can be realized in the presence of
dynamical scalar field, or
it may indicate the breakdown of general relativity (GR) on cosmological
scales. The latter case requires a consistent model of gravity that
explains the accelerating expansion on large scales with the
modification of gravity, while neatly evading the stringent constraints
on the deviation from GR at solar system scales
(e.g., \cite{Khoury:2003rn,Deffayet:2001uk,Hu:2007nk,Starobinsky:2007hu}).
In this respect, the large-scale structure offers the best opportunity
to distinguish between modified gravity and dark energy models in GR, and
the measurement of RSD is a powerful tool to probe gravity.

Given the high-precision measurements of RSD in the near future,
accurate theoretical templates of the redshift-space power spectrum or
correlation function is highly demanded in order to detect a small deviation
of gravity from GR. This is indeed now active research subject, and there are
many studies to accurately model RSD. The RSD measurement is basically made
at the scales close to the linear
regime of gravitational evolution, but the nonlinearity arising both from
the gravity and the RSD is known to play a crucial role. Moreover, due
to the non-Gaussian nature of RSD \cite{Scoccimarro:2004tg},
the applicable range of linear theory prediction is fairly narrower than
that in real space. Thus, beyond the linear scales, a sophisticated treatment
is required for reliable theoretical predictions with a wider applicable range.
Recently, based on the perturbation theory of large-scale structure,
several improved models of RSD have been proposed
\cite{Taruya:2010mx,Nishimichi:2011jm,Matsubara:2007wj,Reid:2011ar,Carlson:2012bu,Seljak:2011tx,Wang:2013hwa,Vlah:2012ni,Vlah:2013lia}. These models properly
account for the non-Gaussian nature of RSD, and are tested
against $N$-body simulations, successfully describing redshift-space
power spectrum and/or correlation function at weakly nonlinear regime.
Applying these models to real observations, constraints on the growth
of structure have been also obtained (e.g., \cite{Reid:2012sw,Blake:2011wn}).

However, it should be noted that the proposed models of RSD have been so
far tested only in the case of GR. Further, beyond linear theory,
the template of RSD is computed with the perturbation theory under the
assumption that gravity is described by GR. Thus, the observational
constraints derived from the PT-based template can be only used as a
consistency test with GR, and a care must be taken in addressing the
constraint on a specific model of modified gravity.

The aim of the present paper is to examine these issues based on an improved
model of RSD developed by Ref.~\cite{Taruya:2010mx}. The power spectrum
expression of this model is similar to the one proposed by
Ref.~\cite{Scoccimarro:2004tg} and
the so-called streaming model frequently used in the literature
(e.g., \cite{Peebles:1980,Fisher:1994ks,Hatton:1997xs,
Scoccimarro:2004tg,Jennings:2010ne}), but it includes two important PT
corrections as a result of the
low-$k$ expansion. Although the model also includes a
phenomenological term to account for the Finger-of-God damping arising from
the small-scale physics, combining the recently developed resummed PT,
it successfully describes not only the matter but also the
halo power spectra in $N$-body simulations
\cite{Taruya:2013my,Nishimichi:2011jm,Ishikawa:2013aea}.
It is shown that the model can be used as a theoretical template to
simultaneously constrain the parameters associated with the cosmic expansion
and the structure growth in an unbiased manner, and
applying it to the Luminous Red Galaxy sample of Sloan Digital Sky Survey
Data Release 7, a robust contraint is obtained 
\cite{Oka:2013cba}.

Here, extending these previous works in GR,
we put forward a prescription to compute the redshift-space power
spectrum in modified gravity models. As a specific example, we explicitly
compute the redshift-space power spectrum in $f(R)$ gravity model, as
one of the representative modified theory of gravity 
\cite{Hu:2007nk,Starobinsky:2007hu}.
The theoretical prediction based on the
standard PT calculation is compared with the results of $N$-body simulations,
and a good agreement is found. Then, we will discuss the potential impact
of the precision modeling of RSD on the model-independent test of GR
and/or constraint on modified gravity models. We will show that
a tight and unbiased constraint on modified gravity models is achieved
only with an improved PT model of RSD
in which the effect of modified gravity is properly taken into account
in the PT calculation. With the improved PT template, testing GR will
be made possible in a model-independent way, but we argue that
a quantitative characterization of the small deviation from GR
generally requires a prior knowledge of modified gravity models.

The paper is organized as follows. In Sec.~\ref{sec:RSD_model}, we begin
by briefly reviewing the model of RSD proposed by Ref.~\cite{Taruya:2010mx}.
Employing the standard PT calculation,
we then give a prescription on how to compute the redshift-space power
spectrum in modified gravity models. In Appendix \ref{sec:BasicEqs_MG},
we summarize the basic formalism of the standard PT in a general
context of modified gravity models, and explicitly give expressions for
the second-order PT kernels used to compute the higher-order corrections of
RSD. In Sec.~\ref{sec:RSD_fRmodel}, as one of the representative models of
modified gravity, we consider the $f(R)$ gravity model, and quantitatively
compare the PT predictions in redshift space with $N$-body
simulations. Based on this, in Sec.~\ref{sec:implications}, a potential impact
of the precision PT model of RSD is discussed, particularly focusing on a
precision constraint on the model parameter of modified gravity, and
model-independent analysis of detecting or characterizing a small deviation
from GR. Finally, Sec.~\ref{sec:conclusion} is devoted to the summary
and conclusion.


\section{Modeling redshift-space power spectrum from perturbation theory}
\label{sec:RSD_model}

\subsection{An improved model of RSD}

We begin by writing the exact expression for redshift-space power spectrum.
Let us denote the density and velocity fields by $\delta$ and $\bfv$.
Owing to the distant-observer approximation, which is usually valid for
the observation of distant galaxies of our interest, one can write
(e.g., \cite{Scoccimarro:2004tg,Bernardeau:2001qr,Taruya:2010mx})
\begin{align}
&P^{\rm(S)}(\bfk)=\int d^3\bfx\,e^{i\bfk\cdot\bfx}
\bigl\langle e^{ik\mu\, \Delta u_z}
\nonumber\\
&\qquad\quad\times
\left\{\delta(\bfr)-\nabla_zu_z(\bfr)\right\}
\left\{\delta(\bfr')-\nabla_zu_z(\bfr')\right\}\bigr\rangle,
\label{eq:Pkred_exact}
\end{align}
where $\bfx=\bfr-\bfr'$ denotes the separation in real space
and $\langle\cdots\rangle$ indicates an ensemble average.
In the above expression, the $z$-axis is taken as
an observer's line-of-sight direction, and we define the directional cosine
$\mu$ by $\mu=k_z/k$. Further,
we defined $u_z(\bfr)=v_z(\bfr)/(aH)$, and $\Delta u_z=u_z(\bfr)-u_z(\bfr')$
for the line-of-sight component of the velocity field.
Note that the above expression has been derived
without invoking the dynamical information
for velocity and density fields, i.e., the Euler equation and/or continuity
equations. Thus, Eq.~(\ref{eq:Pkred_exact}) does hold even in modified gravity
models.

Eq.~(\ref{eq:Pkred_exact}) can be re-expressed in terms of the
cumulants. Then, the term in the bracket is factorized into two
terms, each of which includes the exponential factor (e.g.,
see Eq.(6) of Ref.~\cite{Taruya:2010mx} for explicit expression). Among these,
the overall factor, expressed as $\exp\{\langle e^{ik\mu\,\Delta u_z}\rangle_c\}$
with $\langle\cdots\rangle_c$ being the cumulant,
is responsible for the suppression of the power spectrum arising
mostly from the virialized random and coherent motion on small scales.
The effect of this is known to be partly non-perturbative, and seems
difficult to treat petrubatively. Since it has been shown
to mainly change the broadband shape of the power spectrum, we may
phenomenologically characterize it with a general functional form
$D_{\rm FoG}(k\mu\sigmav)$ with $\sigmav$ being a scale-independent
constant. On the other hand, the remaining factor
includes the term leading to the Kaiser effect in the linear regime
\cite{Kaiser:1987qv,1992ApJ385L5H}, and is likely to affect the
structure of power spectrum on large-scales. Although there also appears
the exponential factor $e^{ik\mu\,\Delta u_z}$ in each term of this factor,
these contributions should be small as long as we consider the large scales,
and the perturbative treatment may be applied.

With the proposition given above, Ref.~\cite{Taruya:2010mx} applied the
low-$k$ expansion, keeping the overall prefactor as general functional
form $D_{\rm FoG}$. The resultant power spectrum expression at one-loop order
becomes
\begin{align}
&P^{\rm(S)}(k,\mu)=\DFoG[k\mu\,\sigmav]\,
\nonumber\\
&\quad\times\,\Bigl\{
P_{\rm Kaiser}(k,\mu)+A(k,\mu)+B(k,\mu)
\Bigr\}, 
\label{eq:TNS_model}
\end{align}
which we hereafter call TNS model.  
Owing to the single-stream approximation in which the
dynamics of large-scale structure is described by the density $\delta$
and velocity divergence $\theta=\nabla\cdot\bfv/(aH)$, the quantities
$P_{\rm Kaiser}$, $A$, and $B$ are explicitly written as
\begin{align}
&P_{\rm Kaiser}(k,\mu)=\Pdd(k)-2\,\mu^2\,\Pdv(k)+\mu^4\,\Pvv(k),
\label{eq:Kaiser}
\\
&A(k,\mu)= -k\mu\,\int \frac{d^3\bfp}{(2\pi)^3} \,\,\frac{p_z}{p^2}
\nonumber\\
&\qquad\quad\times
\left\{B_\sigma(\bfp,\bfk-\bfp,-\bfk)-B_\sigma(\bfp,\bfk,-\bfk-\bfp)\right\},
\label{eq:A_term}
\\
&B(k,\mu)= (k\mu)^2\int \frac{d^3\bfp}{(2\pi)^3} F(\bfp)F(\bfk-\bfp)\,\,;
\label{eq:B_term}
\\
&\qquad\quad F(\bfp)=\frac{p_z}{p^2}
\left\{ \Pdv(p)-\frac{p_z^2}{p^2}\,\Pvv(p)\,\right\},
\nonumber
\end{align}
where $\Pdd$, $\Pvv$, and $\Pdv$ respectively denote auto-power spectra
of density and velocity divergence, and their cross power spectrum.
The function $B_\sigma$ is cross bispectrum defined by
\begin{align}
&\left\langle \theta(\bfk_1)
\left\{\delta(\bfk_2)-\,\frac{k_{2z}^2}{k_2^2}\theta(\bfk_2)\right\}
\left\{\delta(\bfk_3)-\,\frac{k_{3z}^2}{k_3^2}\theta(\bfk_3)\right\}
\right\rangle
\nonumber\\
&\quad\qquad
=(2\pi)^3\delta_D(\bfk_1+\bfk_2+\bfk_3)\,B_\sigma(\bfk_1,\bfk_2,\bfk_3).
\label{eq:def_B_sigma}
\end{align}
Note that in deriving Eq.~(\ref{eq:TNS_model}),
we do not assume any gravity model.
Although the expression (\ref{eq:TNS_model}) has been originally derived
based on the consideration in GR,  as long as the deviation from GR is
small, Eq.~(\ref{eq:TNS_model}) can apply to any model of modified gravity.

As we mentioned in Sec.~\ref{sec:intro}, the main characteristic of the model
given in Eq.~(\ref{eq:TNS_model}) is  the two additional terms $A$ and $B$,
which represent the higher-order coupling between velocity and density fields.
It has been shown in previous studies in GR that these two terms enhance
the power spectrum amplitude over
the scales of baryon acoustic oscillations, 
and moderately but notably change the
acoustic structure imprinted in the power spectrum \cite{Taruya:2010mx}.
As a result, the model (\ref{eq:TNS_model}) successfully describes both the
matter and halo power spectra of $N$-body simulations
at weakly nonlinear scales
\cite{Taruya:2010mx,Nishimichi:2011jm,Taruya:2013my}.
These features are expected to hold qualitatively even in the modified
theory of gravity, but quantitative aspect of the RSD would generally
differ from that of GR, which we will study in detail.

\subsection{Perturbation theory treatment}

To compute the redshift-space power spectrum beyond linear theory,
we apply the PT treatment of gravitational evolution, and
calculate each term in
Eq.~(\ref{eq:TNS_model}) in the quasilinear regime. While the power
spectrum calculation in the case of
GR has been made possible with resummed PT scheme up to the two-loop order
(e.g., \cite{Taruya:2013my} in redshift space, and \cite{Taruya:2009ir,Okamura:2011nu,Crocce:2007dt,Crocce:2012fa,Taruya:2012ut} in real space)
and the applicable range of
the PT prediction has become wider,
we here work with the standard PT calculation at one-loop order for
the predictions in modified theory of gravity. Although the standard PT
treatment in GR is known to produce ill-behaved PT expansion that lacks
good convergence properties (e.g., \cite{Crocce:2005xy,Carlson:2009it,
Taruya:2009ir,Bernardeau:2012ux}),
using the standard PT as theoretical template,
we can still get a fruitful cosmological constraint at the quasilinear
scales (e.g., \cite{Saito:2010pw,Zhao:2012xw}).
In the present paper, we use
the standard PT formalism developed by Ref.~\cite{Koyama:2009me}, which is
suited to deal with a wide class of modified gravity models.
In what follows, we separately give a prescription on how to compute
the power spectrum corrections in Eq.~(\ref{eq:TNS_model}).

\subsubsection{Non-linear Kaiser term}

The term $P_{\rm Kaiser}$ in Eq.~(\ref{eq:TNS_model}) is
the leading-order contribution to the redshift-space power spectrum.
In the large-scale limit where the linear theory prediction is safely applied,
we have $\theta=-f\,\delta$, and Eq.~(\ref{eq:Kaiser}) is reduced to
the Kaiser formula, $P_{\rm Kaiser}=(1+f\,\mu^2)^2\,\Pdd(k)$ \cite{Kaiser:1987qv},
where $f$ is the linear growth rate defined by $d\ln\,D_+/d\ln a$
with $D_+$ being the linear growth factor. Beyond the linear theory,
a simple relation between density and velocity divergence fields
no longer hold, and we need to separately evaluate the three power spectra,
$\Pdd$, $\Pdv$ and $\Pvv$, especially in the modified gravity models
\cite{Scoccimarro:2004tg}.

In contrast to the GR, one crucial point in the modified gravity models is
that a new scalar degree of freedom,
sometimes referred to as the scalaron, arises and modifies the
force law. In the presence of an extra scalar
field, even though the conservation law of energy momentum tensor remains
unchanged, the Poisson equation is inevitably modified and is coupled
to the field equation for scalaron. In particular,
in successful modified gravity models that have a mechanism to recover
GR on small scales, the scalaron generally
acquires nonlinear interaction terms, and they play
an important role to recover GR on small scales. Thus, we need to
properly take into account such a nonlinear
interaction of the scalaron, and consistently
solve the evolution equations of density and velocity divergence.

In Ref.~\cite{Koyama:2009me}, we have developed a
formalism to calculate the nonlinear power spectrum in a
wide class of modified gravity models, including $f(R)$ gravity and
Dvali-Gabadadze-Porratti (DGP) braneworld \cite{Dvali:2000hr} models.
The formalism perturbatively treats the effect of
nonlinear scalaron, and employing the standard PT technique,
we have explicitly
computed the power spectrum of density field, $\Pdd$, at one-loop order,
which reproduces the $N$-body results at quasi-linear scales.
In what follows, we adopt this formalism to perturbatively
compute auto- and cross-power spectra of density and velocity divergence.
The basic equations for perturbations are briefly summarized in Appendix
\ref{sec:BasicEqs_MG}.

To compute the one-loop power spectra, specifically in the $f(R)$ gravity
model presented below (Sec.~\ref{sec:RSD_fRmodel}),  the analytic calculations
starting naively with the basic equations in Appendix \ref{sec:BasicEqs_MG}
is technically difficult in practice, because the
perturbation equations cannot be separately treated in time and scales.
Instead of solving the equations for $\delta$ and $\theta$
in Appendix \ref{sec:BasicEqs_MG}, we will numerically solve the evolution
equations for the power spectra $P_{ab}$ called the closure equation
in Ref.~\cite{Koyama:2009me} [Eqs.~(4.3)(4.4)(4.5)],
which has been derived by truncating an infinite chain of the moment
equations at one-loop order. Implementation and technical detail
of the numerical scheme to solve the closure equations are presented
in Ref.~\cite{Hiramatsu:2009ki}
(see also Appendix A of Ref.~\cite{Koyama:2009me}).

\subsubsection{$A$ term}

Next consider the $A$ term.
The expression given in Eq.~(\ref{eq:A_term}) can be rewritten with a
more convenient form suited for numerical integration.
Introducing the doublet\footnote{This is
somewhat different from the frequently-used definition in GR,
$\Phi_a=(\delta,-\theta/f)$ with $f$ being the linear growth rate,
$f\equiv d\ln D_+/d\ln a$ (e.g., \cite{Bernardeau:2008fa,Taruya:2009ir}).}, 
$\Phi_a\equiv(\delta,\theta)$, we define the bispectrum $B_{abc}$:
\begin{align}
& \langle\Phi_a(\bfk_1)\Phi_b(\bfk_2)\Phi_c(\bfk_3)\rangle
\nonumber\\
&\qquad=(2\pi)^3\,\delta_D(\bfk_1+\bfk_2+\bfk_3)\,B_{abc}(\bfk_1,\bfk_2,\bfk_3).
\end{align}
In terms of this, the three-dimensional integral is reduced to the
sum of the two-dimensional integrals, and the final form of the $A$ term
becomes \cite{Taruya:2013my}
\begin{align}
& A(k,\mu)
=\sum_{n=1}^3\sum_{a,b}^2\,\mu^{2n}\,(-1)^{a+b-1}\,
\frac{k^3}{(2\pi)^2}\,
\nonumber\\
&\times
\int_0^\infty dr\int_{-1}^{1} dx
\,\bigl\{A^n_{ab}(r,x)\,B_{2ab}(\bfp,\bfk-\bfp,-\bfk)
\nonumber\\
&\qquad\qquad+\widetilde{A}^n_{ab}(r,x)\,B_{2ab}(\bfk-\bfp,\bfp,-\bfk)
\bigr\}
\label{eq:A_term_simplified}
\end{align}
with $r=p/k$ and $x=\bfk\cdot\bfp/(kp)$.
The non-vanishing components of $A^a_{bc}$ and $\widetilde{A}^a_{bc}$
are exactly the same as those presented in Ref.~\cite{Taruya:2013my}
(see Sec.~III-B2).

Since the $A$ term appears as a next-to-leading order correction,
the tree-level calculation of the bispectrum is sufficient
for a consistent calculation of redshift-space power spectrum at one-loop
order. Expanding the doublet $\Phi_a$ as
$\Phi_a=\Phi_a^{(1)}+\Phi_a^{(2)}+\cdots$, and assuming the Gaussian initial
condition, the bispectrum at the tree-level order becomes
\begin{align}
&B_{abc}(\bfk_1,\bfk_2,\bfk_3;t)
\nonumber\\
&\quad\,=2\Big\{F_a^{(2)}(\bfk_2,\bfk_3;t)F_b^{(1)}(k_2;t)F_c^{(1)}(k_3;t)
\nonumber\\
&\qquad\qquad\qquad\quad
\times P_0(k_2)P_0(k_3)\,+\,\,\mathrm{(cyc. perm.)}\,\,
\Big\},
\end{align}
where the functions $F_a^{(n)}$ are the symmetrized
standard PT kernel of the $n$-th order
perturbative solutions\footnote{Since we are
interested in the late-time evolution of cosmic structure,
we only consider the fastest growing term.}:
\begin{align}
& \Phi_a^{(n)}(\bfk;t) = \int\frac{d^3\bfk_1 \cdots d^3\bfk_n}{(2\pi)^{3(n-1)}}
\,\delta_D(\bfk-\bfk_{1\cdots n})\,
\nonumber\\
&\qquad\qquad\,\,\times
F_a^{(n)}(\bfk_1,\cdots,\bfk_n;t) \,\delta_0(\bfk_1)\cdots\delta_0(\bfk_n)
\label{eq:PT_kernel}
\end{align}
with $\bfk_{1\cdots n}=\bfk_1+\cdots+\bfk_n$.
The function $\delta_0$ is the initial density field, for which
we assume Gaussian statistics. The statistical
property of $\delta_0$ is characterized by the power spectrum:
\begin{align}
\langle\delta_0(\bfk)\delta_0(\bfk')\rangle=
(2\pi)^3\,\delta_D(\bfk_1+\bfk')\,P_0(k)
\label{eq:initial_pk}
\end{align}

The remaining task in computing the $A$ term
is to evaluate the PT kernels up to the second order,
which can be done analytically. This is also the case
with the $f(R)$ gravity model given below, although some numerical works
are involved. In Appendix \ref{sec:BasicEqs_MG},
based on the basic equations,
we derive the explicit functional form of the PT kernels $F_a^{(1)}$ and
$F_a^{(2)}$, and summarize the procedure to compute these kernels.
The explicit expression for the
PT kernels in $f(R)$ gravity and DGP models is also presented.

\subsubsection{$B$ term}

Similar to the $A$ term, the $B$ term given in Eq.~(\ref{eq:B_term}) is
reduced to the sum of two-dimensional integrals. From
Refs.~\cite{Taruya:2010mx,Taruya:2013my}, the resultant expression
becomes
\begin{align}
&B(k,\mu)=\sum_{n=1}^4\sum_{a,b=1}^2 \mu^{2n}\frac{k^3}{(2\pi)^2}\,
\int_0^\infty dr \int_{-1}^1\,dx\,\,
\nonumber\\
&\qquad\qquad\quad
\times B_{ab}^n(r,x)\,\frac{P_{a2}(k\sqrt{1+r^2-2rx})\,P_{b2}(kr)}{(1+r^2-2rx)^a},
\label{eq:B_term2}
\end{align}
Here, the coefficients $B_{ab}^n$ are the same as presented in Appendix B
of Ref.~\cite{Taruya:2010mx} \footnote{In the case of GR,
Eq.~(\ref{eq:B_term2}) exactly
coincides with Eq.~(A4) of Ref.~\cite{Taruya:2010mx}, but
the definition of power spectra $P_{ab}$ is somewhat different. }.
For the one-loop order of the redshift-space power spectrum,
the linear-order power spectra are sufficient to evaluate
Eq.~(\ref{eq:B_term2}), and we have
\begin{align}
P_{ab}(k;t)=F_a^{(1)}(k;t)F_b^{(1)}(k;t) P_0(k)
\end{align}
with the linear PT kernel $F_a^{(1)}$.

\section{Redshift-space distortions in $f(R)$ gravity model}
\label{sec:RSD_fRmodel}

In this section, as an illustrative example showing the RSD beyond linear
scales in modified gravity model,
we compute the redshift-space power spectrum in the $f(R)$
gravity model, and compare the PT prediction with results of
$N$-body simulations.

\subsection{$f(R)$ gravity}

The $f(R)$ gravity is one of the representative gravity models,
and it has a mechanism
to recover GR on small scales. Generalizing the
Einstein-Hilbert action to include an arbitrary function of
the scalar curvature $R$, the model is given by:
\begin{equation}
S=\int d^4x \sqrt{-g}\left[\frac{R+f(R)}{2 \kappa^2} +L_{\rm m}\right],
\label{eq:action_fR}
\end{equation}
where $\kappa^2=8\pi\,G$ and $L_{\rm m}$ is the Lagrangian of the ordinary
matter. This theory is equivalent to the Brans-Dicke theory
with parameter $\omega_{\rm BD}=0$, but
there is a non-trivial potential. 
This can be seen from the trace of modified Einstein equations:
\begin{equation}
3 \Box f_R - R + f_R R -2 f = - \kappa^2 \rho,
\end{equation}
where $f_R = df/dR$ and $\Box$ is a Laplacian operator and we assumed
the matter dominated universe. 
We can identify $f_R$ as the scalaron, i.e., extra scalar field, and its
perturbations are defined as
\begin{equation}
\varphi = \delta f_R\equiv f_R-\overline{f}_R,
\label{eq:scalaron_fR}
\end{equation}
where the bar indicates that the quantity is evaluated on the
background universe. Here, we consider the cases with $| \bar{f}_R | \ll 1$
and $|\bar{f}/\bar{R}| \ll 1$. These conditions are necessary to have the background close to cold dark matter ($\Lambda$CDM) cosmology. Then the perturbations for scalaron
satisfy
\begin{equation}
3 \frac{1}{a^2} \nabla^2 \varphi = - \kappa^2 \rho_m \delta
+ \delta R,
\quad \delta R \equiv R(f_R)-R(\overline{f}_R).
\end{equation}
Note that this is nothing but the equation for the BD scalar perturbations
with $\omega_{\rm BD}=0$.

In what follows, we consider the specific function $f(R)$ of the form
\begin{equation}
f(R) \propto \frac{R}{A R +1},
\label{eq:model}
\end{equation}
where $A$ is a constant with dimensions of length squared \cite{Hu:2007nk}.
If we take the limit $R \to 0$, we obtain $f(R) \to 0$ and 
cosmological constant does not appear. 
For high curvature $A R \gg 1$, on the other hand, $f(R)$ can be expanded as
\begin{equation}
f(R) \simeq - 2 \kappa^2 \rho_{\Lambda} + |f_{R0}|\, \frac{\bar{R}_0^2}{R},
\label{eq:model_approx}
\end{equation}
where $\rho_{\Lambda}$ is determined by $A$. The quantity $\bar{R}_0$
is the background curvature today, and we defined 
$f_{R0}=\bar{f}_R(R_0)$ (see e.g., \cite{Marchini:2013oya,Lombriser:2010mp,Yamamoto:2010ie,Okada:2012mn,Schmidt:2009am} 
for recent cosmological constraints on $|f_{R0}|$).

In the setup of $N$-body simulations and PT calculation below, we will 
take $|f_{R0}| \ll 1$, and assume that the background expansion just 
follows the $\Lambda$CDM history with the same $\rho_{\Lambda}$.

\subsection{$N$-body simulations}

We use the subset of cosmological $N$-body simulations
presented in Ref.~\cite{Li:2012by, Jennings:2012pt}. The data set of $N$-body simulations
were created by
the $N$-body code for modified gravity models, {\tt ECOSMOG} \cite{Li:2011vk},
which is a modification of the mesh-based N-body code, {\tt RAMSES}
\cite{Teyssier:2001cp}.
With cubic boxes of side length $1.5\,h^{-1}\,$Gpc and $1024$ particles,
the initial conditions were generated at redshift $z_{\rm init}=49$
using {\tt MPgrafic}\footnote{\tt{http://www2.iap.fr/users/pichon/mpgrafic.html}}, according to the linear matter power spectrum determined by
the cosmological parameters: $\Omega_{\rm m}=0.24$, $\Omega_{\Lambda}=0.76$,
$\Omega_{\rm b}=0.0481$, $h=0.73$, $n_{\rm s}=0.961$, $\sigma_8=0.801$ \footnote{The value of $\sigma_8$ indicated in Ref.~\cite{Jennings:2012pt} is a typo.}.

In the analysis presented below, we mainly consider
GR and $f(R)$ gravity with $|f_{R0}|=10^{-4}$ (labeled as $\Lambda$CDM
and F4 in Ref.~\cite{Jennings:2012pt}), and
focus on the output results at $z=1$.
In Ref.~\cite{Jennings:2012pt}, with $6$ independent realizations,
the matter power spectra are measured in redshift space,
applying the distant-observer approximation.
Adopting the line-of-sight direction perpendicular to
each side of the simulation box, the density field is assigned on
$1024^3$ grids with cloud-in-cell interpolation, and
the power spectra are measured for three different line-of-sight directions.
Here, in comparison with the PT model, we use the power spectra averaged
over all line-of-sight directions and realizations. With this treatment,
the measured power spectra tend to be smooth and the outliers disappear,
while the estimation of the error covariance is rather complex
because of the duplicated power spectrum measurement with same realizations.
In what follows,
for the analysis of the fitting and parameter estimation,
we assume a hypothetical galaxy survey of the
volume $V=10\,h^3\,$Gpc$^3$,
and consider the statistical error limited by the
cosmic variance. Unless otherwise stated, the error bars of the $N$-body
results indicate the 1-$\sigma$ error of this hypothetical survey
computed with the linear power
spectrum (see Appendix C of Ref.~\cite{Taruya:2010mx})\footnote{Eq.~(C4)
of Ref.~\cite{Taruya:2010mx} includes typos. In the parenthesis of the third
line, it should be correctly replaced with $5+(110/21)\beta+(15/7)\beta^2$. }

Finally, in addition to the redshift-space power spectra, we also compare
the real-space power spectra of the density and velocity fields in
Ref.~\cite{Jennings:2012pt}, which are used to estimate the valid range
of PT predictions.

\subsection{Comparison with $N$-body simulations}

\begin{figure}[t]
\begin{center}
\includegraphics[width=8.5cm,angle=0]{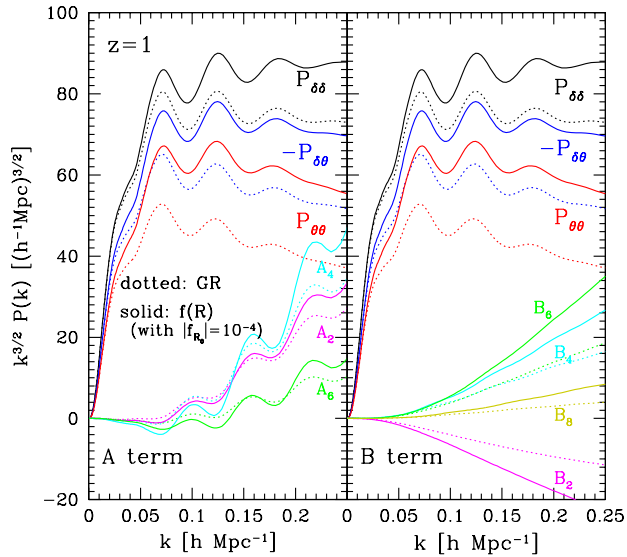}
\end{center}

\vspace*{-1.0cm}

\caption{Power spectrum corrections from $A$ term (left) and
$B$ term (right). The plotted results are at $z=1$, and they are
multiplied by $k^{3/2}$ just for illustrative purpose. The $A$ and $B$
terms are respectively expanded as
$A(k,\mu)=\sum_n^3 A_{2n}(k)\mu^{2n}$ and $B(k,\mu)=\sum_n^4 B_{2n}(k)\mu^{2n}$,
and we here plot the scale-dependent coefficients $A_{2n}$ and $B_{2n}$.
The dotted lines are the results in GR, while the solid lines are those in
$f(R)$ gravity model with $|f_{R_0}|=10^{-4}$. Those results are computed with
the same linear power spectrum (see text).
For reference, the power spectra $P_{\delta\delta}$, $P_{\delta\theta}$, and
$P_{\theta\theta}$ are also shown in black, blue and red lines for GR case.
\label{fig:pkcorr_A_B}}
\end{figure}

Before presenting the redshift-space power spectrum, we
first separately compute the contribution of each term in the power spectrum
expression (\ref{eq:TNS_model}), and compare it with $N$-body
simulation. Fig.~\ref{fig:pkcorr_A_B} shows the results of standard PT
calculation at one-loop order at $z=1$ in the GR (dotted) and F4
($f(R)$ gravity with $|f_{R,0}|=10^{-4}$, solid)
cases, with the same cosmological parameters as adopted in
$N$-body simulations. The $A$ (left) and $B$ (right) terms
are plotted together with the auto- and cross-power spectra of
density and velocity divergence fields, multiplied by $k^{3/2}$. According to
Eqs.~(\ref{eq:A_term_simplified}) and (\ref{eq:B_term2}), the $A$ and $B$
terms are expanded as $A(k,\mu)=\sum_n^3 A_{2n}(k)\mu^{2n}$ and
$B(k,\mu)=\sum_n^4 B_{2n}(k)\mu^{2n}$, and we here plot the
scale-dependent coefficients $A_{2n}$ and $B_{2n}$
($A_2,\,\,B_2$: magenta, $A_4,\,\,B_4$: cyan, $A_6,\,\,B_6$: green, $B_8$:
yellow). Overall, the resultant amplitude of power spectrum corrections in F4
is rather larger than that in GR. The acoustic signature are clearly
seen in both GR and F4 cases not only for the real-space quantities but also
the RSD correction (i.e., $A$ term).
The reason for a larger amplitude in $f(R)$ gravity is mainly attributed to
the scale-dependent enhancement of the linear growth factor on small scales in
$f(R)$ gravity, and with a large value of $|f_{R,0}|=10^{-4}$,
the mechanism to recover GR is still inefficient at quasi-linear scales.
This implies that the redshift-space power spectrum
can be quite different between GR and F4, and the RSD corrections
(i.e., $A$ and $B$ terms) would play an important role.

\begin{figure}[t]
\begin{center}
\includegraphics[width=8.5cm,angle=0]{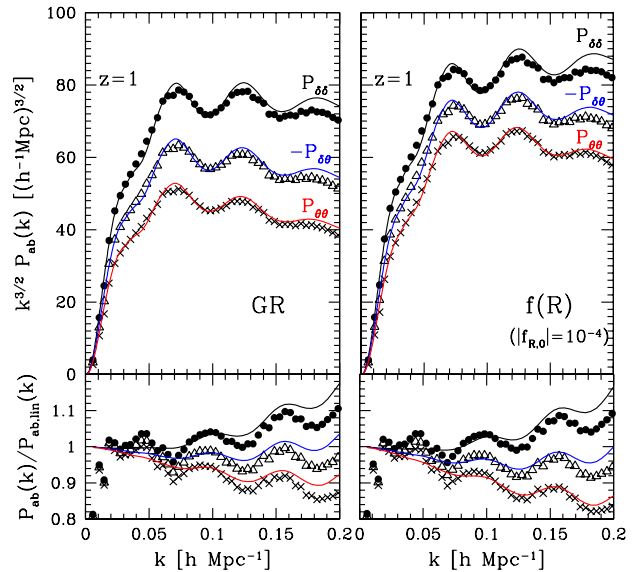}
\end{center}

\vspace*{-0.5cm}

\caption{Auto and cross power spectra of density and velocity fields in real space at $z=1$ for GR (left) and $f(R)$ gravity with $|f_{R,0}|=10^{-4}$.
Top panels shows the power spectra
multiplied by the cube of wavenumber, i.e., $k^{3/2}P_{ab}(k)$, while bottom
panels present the ratio of the power spectra to the linear theory
predictions, $P_{ab}(k)/P_{ab,{\rm lin}}(k)$.
\label{fig:pk_dd_dt_tt}}
\end{figure}

To see the domain of applicability of the standard PT
calculation, we next plot in
Fig.~\ref{fig:pk_dd_dt_tt} the real-space power spectra
$\Pdd$, $\Pdv$, and $\Pvv$ from $N$-body simulations, and compare those
with the PT results. According to a phenomenological rule
calibrated with $N$-body simulations \cite{Nishimichi:2008ry,Taruya:2009ir}, the standard PT power spectrum at
one-loop order is expected to agree with $N$-body simulations at
$k\lesssim0.12\,(0.15)\,\,h\,$Mpc$^{-1}$ with an
accuracy of $1\%\,\,(3\%)$ level. Although the simulation results show
somewhat noisy behavior, the PT predictions seem to work well at least
at the scales indicated by the empirical rule, where the deviation from
linear theory is around $10\%$ in both GR and F4. The result suggests
that even in the presence of a substantial difference in the linear growth,
the nonlinear gravitational growth itself does not change so much between
GR and modified gravity models. This would be probably
true as long as the mechanism to recover GR is still inefficient
at quasi-linear scales.

\begin{figure}[b]
\begin{center}
\includegraphics[width=8.5cm,angle=0]{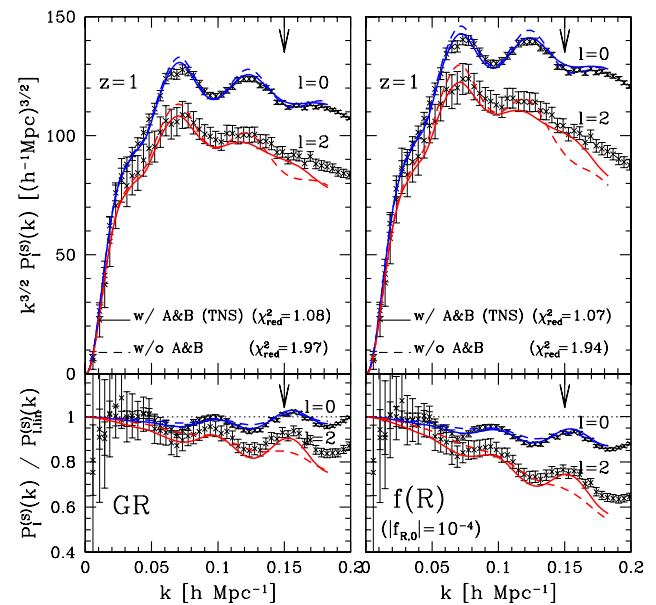}
\end{center}

\vspace*{-0.5cm}

\caption{Monopole (blue) and quadrupole (red) moments of 
redshift-space power spectra
at $z=1$ for GR (left) and $f(R)$ with $|f_{R,0}|=10^{-4}$ (right).
Top panels show the monopole and quadrupole power spectra
multiplied by $k^{3/2}$, while bottom panels present the ratio of power
spectra to linear theory predictions,
$P^{\rm(S)}_{\ell}(k)/P^{\rm(S)}_{\ell,{\rm lin}}(k)$. 
Solid and dashed lines respectively show the PT results 
based on the TNS model [Eq.~(\ref{eq:TNS_model})] 
with and without $A$ and $B$ terms.  In each panel, vertical arrow
indicates the maximum wavenumber used to estimate $\sigmav$.
\label{fig:pkred}}
\end{figure}

Keeping in mind the applicability of PT calculation, we now focus on
the redshift-space power spectrum, and
compare the PT calculations with $N$-body simulations.
In Fig.~\ref{fig:pkred},
top panels show the monopole ($\ell=0$) and
quadrupole ($\ell=2$) moments of power spectrum multiplied by $k^{3/2}$,
while bottom panels present the ratio of monopole and quadrupole spectra
to the linear theory prediction taking only account of the Kaiser effect.
The multipole power spectrum $P^{\rm(S)}_\ell$  is defined by
\begin{align}
&P^{\rm(S)}_\ell(k)=\frac{2\ell+1}{2}\,\int_{-1}^{1}d\mu\,P^{\rm(S)}(k,\,\mu)
\,\mathcal{P}_\ell(\mu),
\end{align}
with $\mathcal{P}_\ell$ being the Legendre polynomials.
The PT results based on an improved model of RSD 
[i.e., TNS model, Eq.~(\ref{eq:TNS_model})] 
are depicted as solid lines, while the results without correction terms
are also shown in dashed lines. In both cases, we adopt
the Gaussian damping function in computing PT predictions:
\begin{align}
D_{\rm FoG}(k\mu\,\sigmav) = \exp\left[-(k\mu\,\sigmav)^2\right].
\end{align}
Here, the velocity dispersion $\sigmav$ is a free parameter and
is determined by fitting the model predictions to the $N$-body results of
monopole and quadrupole spectra up to $k_{\rm max}=0.15\,h\,$Mpc$^{-1}$
(indicated by vertical arrows),  corresponding to the valid range of PT.
Note that we also examined the Lorentzian form, but
the choice of the damping function did not change the results
as long as we consider the applicable range of standard PT one-loop.

\begin{figure}[ht]

\vspace*{-1.5cm}

\begin{center}
\includegraphics[width=8.5cm,angle=0]{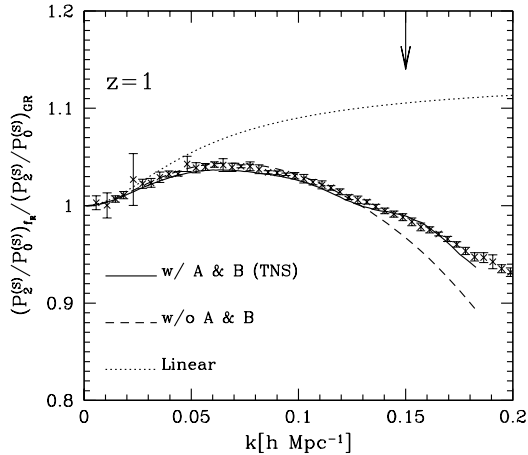}
\end{center}

\vspace*{-1.0cm}

\caption{Ratio of quadrupole-to-monopole ratio of $f(R)$ gravity to that
of GR, $(P_2^{\rm (S)}/P_0^{\rm (S)})_{\rm f(R)}/(P_2^{\rm (S)}/P_0^{\rm (S)})_{\rm GR}$.
The results at $z=1$ is shown. Solid and dashed lines are the PT predictions
based on the TNS model 
with and without $A$ and $B$ terms, while dotted lines are the linear
theory predictions. The vertical arrow
indicates the maximum wavenumber used to estimate $\sigmav$.
\label{fig:QMratio}}
\end{figure}

Fig.~\ref{fig:pkred} shows that
the model (\ref{eq:TNS_model}) successfully describes the $N$-body results of
RSD in both GR and $f(R)$ model. Although
the applicable range of standard PT one-loop is limited,
the $A$ and $B$ terms still play an important role. In the presence
of these terms, the acoustic signature of redshift-space power spectrum
tends to be
smeared compared to the real-space power spectrum, and this
indeed improves the agreement with $N$-body simulations. In the panels of
Fig.~\ref{fig:pkred}, we show the reduced chi-squared statistic defined
by\footnote{Strictly speaking, the non-vanishing monopole and
quadrupole moments of
redshift-space power spectra yield a non-zero covariance between them.
This is true even in the Gaussian statistics.
However, the magnitude of covariance is
shown to be fairly small at large scales \cite{Taruya:2010mx,Taruya:2011tz},
and the impact of covariance is ignorable in our analysis. }
\begin{align}
\chi_{\rm red}^2=\frac{1}{\nu}\sum_{\ell=0,2}\sum_i
\frac{\left[P_{\ell,{\rm N\mbox{-}body}}^{\rm(S)}(k_i)-
P_{\ell,{\rm PT}}^{\rm(S)}(k_i)\right]^2}{[\Delta P^{\rm(S)}_\ell(k_i)]^2},
\label{eq:chi2_red}
\end{align}
with the quantity $\nu$ being the number of degrees of freedom.
Here, the statistical error $\Delta P^{\rm(S)}_\ell$ is estimated
from the cosmic variance error assuming the survey volume
$10\,h^{-3}\,$Gpc$^3$. The number of Fourier bins in the above summation can be
inferred from the maximum wavenumber shown in Fig.~\ref{fig:pkred}, depicted
as vertical arrows. The resultant $\chi_{\rm red}^2$ taking account of
the $A$ and $B$ terms (i.e., TNS model)
are clearly lower than those ignoring the corrections. 

\begin{figure}[ht]
\begin{center}
\includegraphics[width=8.8cm,angle=0]{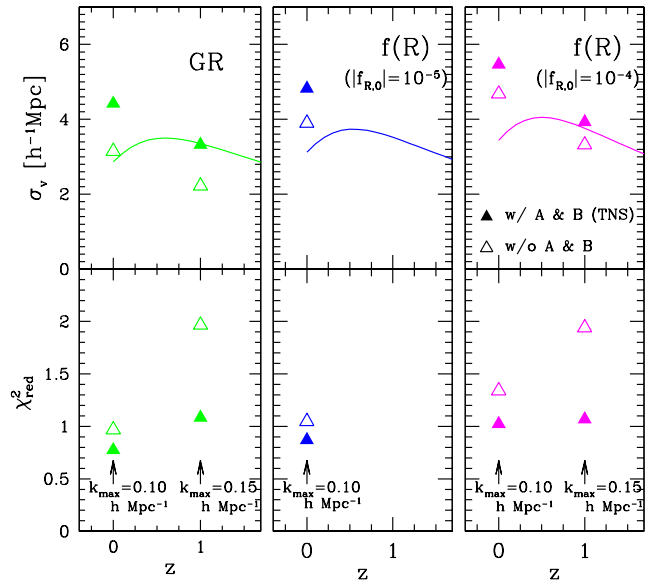}
\end{center}

\vspace*{-0.8cm}

\caption{Fitting results in GR (left) and $f(R)$ gravity with
$|f_{\rm R,0}|=10^{-4}$ (middle) and $10^{-5}$ (right). The best-fit parameter
$\sigma_{\rm v}$ and the resultant values of reduced chi-squared are respectively
shown in top and bottom panels. For comparison, linear theory prediction of
the velocity dispersion,
$\sigma_{\rm v,lin}^2\equiv\int dq P_{\theta\theta,{\rm lin}}(q)/(6\pi^2)$, is also
depicted
as solid lines in top panels. In each panel, filled and open triangles indicate
the PT results based on the TNS model 
with and without $A$ and $B$ terms, respectively. Note that the
maximum wavenumber in the fitting, $k_{\rm max}$, is set to $0.10$ and $0.15\,h$\,Mpc$^{-1}$
at $z=0$ and $1$.
\label{fig:fitting_sigmav2}}
\end{figure}
To show the quantitative difference of RSD between GR and $f(R)$ gravity,
Fig.~\ref{fig:QMratio} shows the ratio of quadrupole-to-monopole ratio
in F4 to that in GR, i.e.,
$(P_2^{\rm(S)}/P_0^{\rm(S)})_{f(R)}/(P_2^{\rm(S)}/P_0^{\rm(S)})_{\rm GR}$. Note that
the errorbars of the $N$-body simulation shown in the panel are not
the cosmic variance error, but are estimated
from the $N$-body data of the $6$ realizations
for a particular line-of-sight direction.
The linear theory predicts a slight enhancement of the ratio, while the
actual $N$-body result rather shows a noticeable reduction at small scales.
This basically comes from a stronger suppression of the
power spectra in $f(R)$ gravity, as shown in Fig.~\ref{fig:pkred} (see
bottom panel). Fig.~\ref{fig:fitting_sigmav2} summarizes the fitting
results of the parameter $\sigmav$ together with the resultant
reduced chi-squared. At $z=1$, the fitted value of the velocity dispersion
is relatively large in F4 by $\sim20\%$. Neglecting the correction terms, the
relative difference of $\sigmav$ between $f(R)$ model and GR is more
prominent ($\sim50\%$), although the values themselves are even smaller
than those taking account of the $A$ and $B$ terms. As a result, the
PT prediction ignoring the corrections exhibits a strong damping behavior in
Fig.~\ref{fig:QMratio},
and tends to deviate from $N$-body simulations at small scales.
By contrast, the
prediction with $A$ and $B$ terms (i.e., TNS model) 
faithfully traces the $N$-body trend
beyond the applicable range of the standard PT.

Finally, while we mainly presented the results at $z=1$, we
briefly comment on other cases at $z=0$, where
we also examined the F5 case ($f(R)$ gravity with $|f_{R,0}|=10^{-5}$).
All the results are summarized in Fig.~\ref{fig:fitting_sigmav2}.
At $z=0$, the nonlinear clustering is strongly developed, and the
applicable range of standard PT one-loop is quite limited. Nevertheless,
with a limited fitting range of $k\leq k_{\rm max}=0.1\,h\,$Mpc$^{-1}$,
the PT results show an excellent performance with $\chi_{\rm red}^2\sim1$,
and the prediction including the $A$ and $B$ terms gives a better agreement
with $N$-body simulations. Fig.~\ref{fig:fitting_sigmav2} shows that
the fitted value of velocity dispersion in
$f(R)$ gravity is generally larger than that in GR, roughly consistent with
the one estimated with linear theory (solid lines). This suggests that
a stronger damping of the power spectrum amplitude may be a good indicator
for modified gravity, as pointed out by
Ref.~\cite{Jennings:2012pt} (see also Ref.~\cite{Lam:2012by,Lam:2013kma}).
Note, however, that the actual value of
$\sigmav$ depends on the underlying model of RSD. Further,
our observable is not dark matter but galaxy distribution, which
does not faithfully trace the dark matter distribution. A careful study is
needed, and we leave this issue to future work.

\begin{figure}[t]
\begin{center}
\includegraphics[width=9cm]{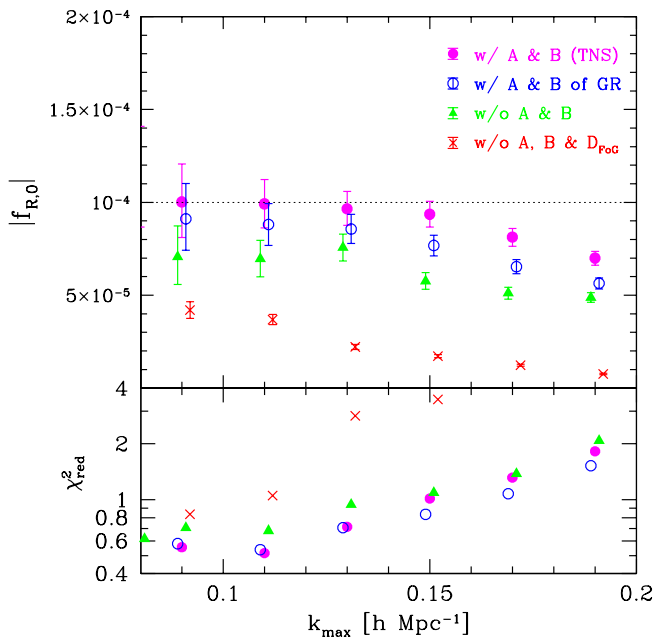}
\end{center}

\vspace*{-0.5cm}

\caption{Top: Best-fit values of $|f_{R,0}|$
as function of the maximum wavenumber $k_{\rm max}$ used for MCMC analysis.
Assuming the cosmic variance limited
survey of the volume $V=10\,h^{-3}\,$Gpc$^3$,
we fit the PT template to the $N$-body
simulation of the F4 run at $z=1$, and derive the best-fit values and
1-$\sigma$ statistical error of $|f_{R,0}|$,
allowing the parameter $\sigmav$ to be free.
Filled circles are
the results based on the TNS model [Eq.~(\ref{eq:TNS_model})] 
in $f(R)$ gravity, 
while filled triangles are the cases ignoring the $A$ and $B$ terms. Open
circles represent the results similar to filled circles, but the corrections
$A$ and $B$ are calculated in GR. For comparison, crosses are the results
ignoring not only the $A$ and $B$ terms but also the damping function
$D_{\rm FoG}$.
\label{fig:fR0_MCMC}}
\end{figure}

\begin{figure*}[tb]
\begin{center}
\hspace*{-0.5cm}
\includegraphics[width=9.5cm]{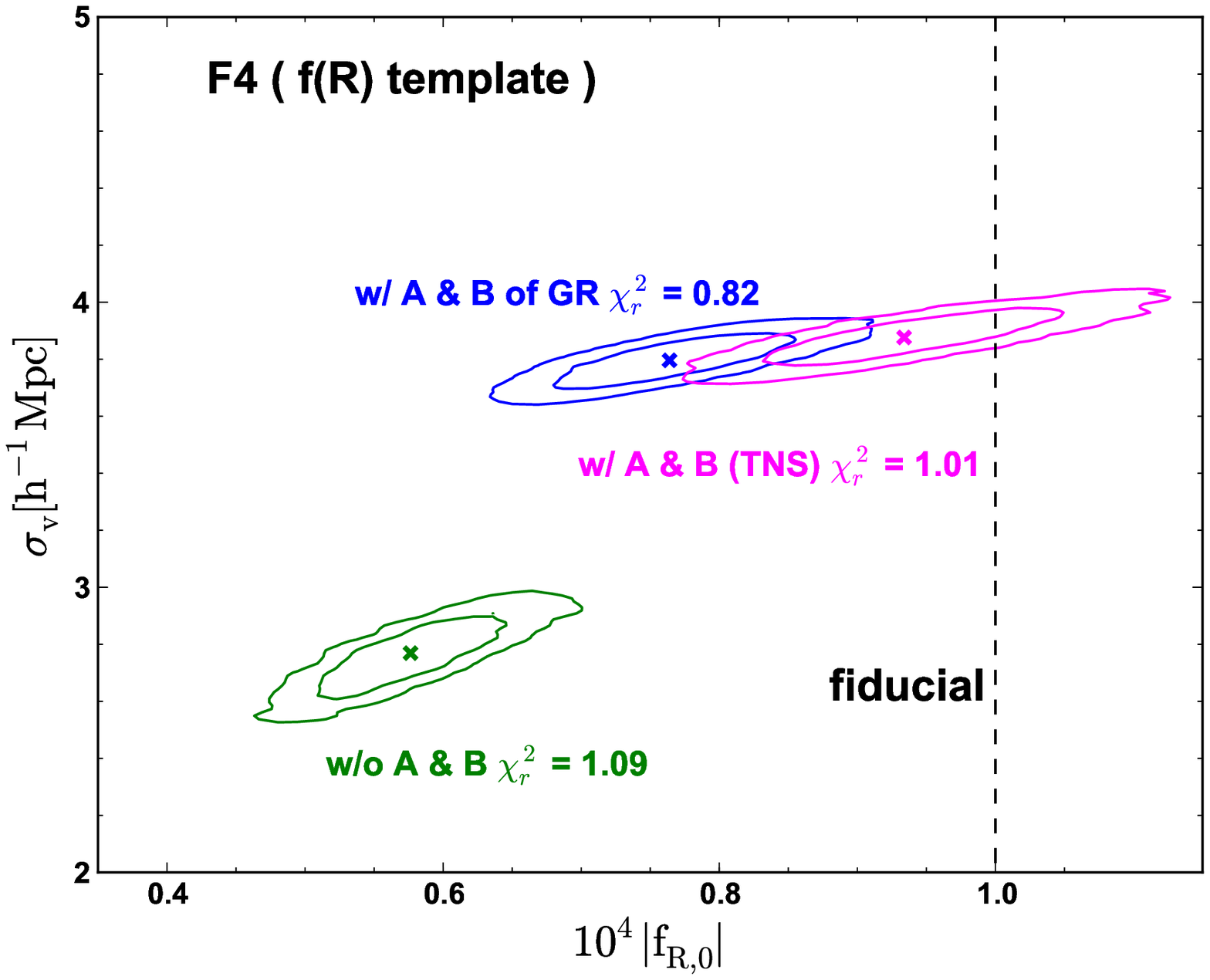}
\hspace*{-1.0cm}
\includegraphics[width=9.5cm,angle=0]{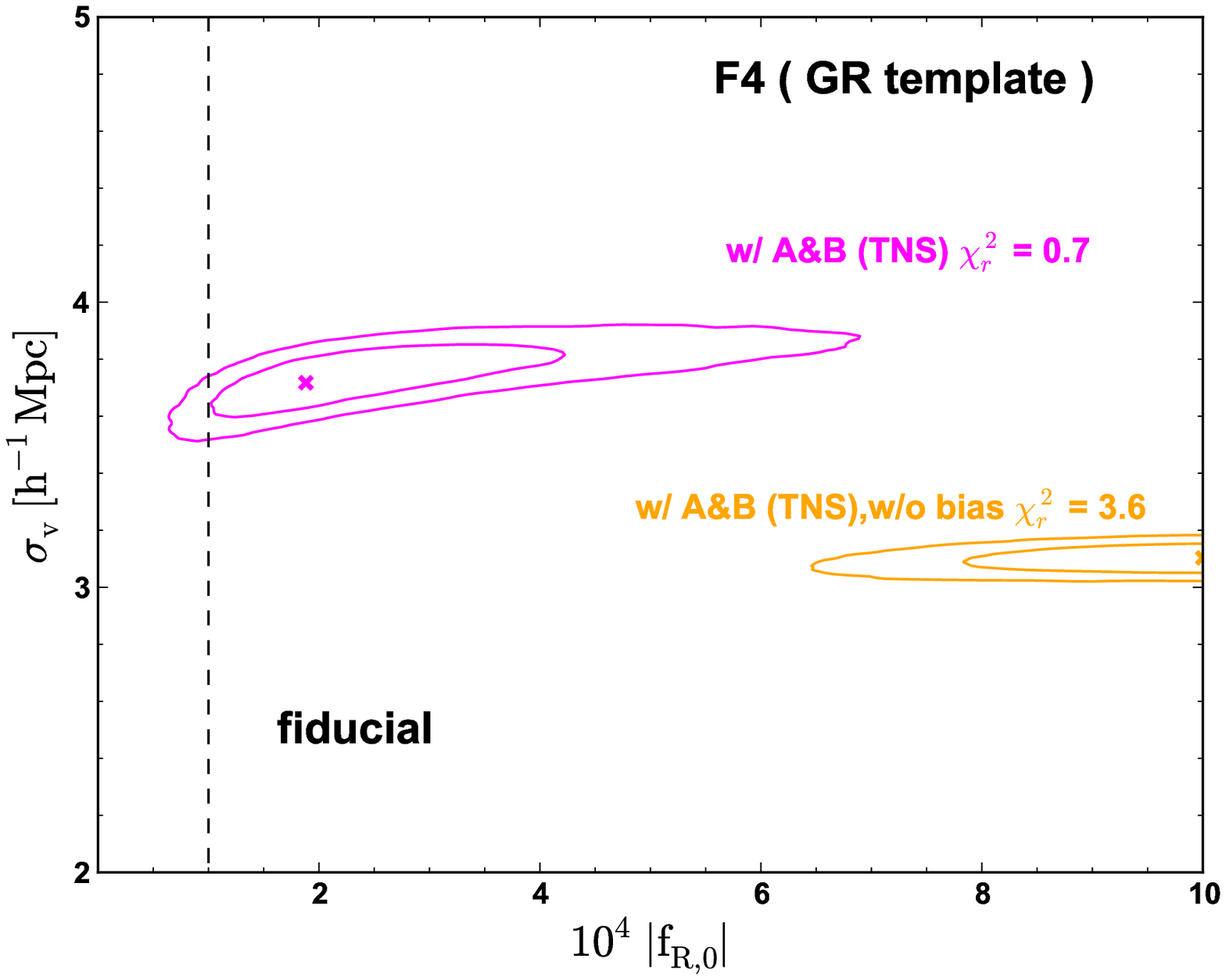}
\end{center}

\vspace*{-0.5cm}

\caption{Two-dimensional error contours derived from MCMC analysis,
fixing the maximum wavenumber to $k_{\rm max}=0.15\,h\,$Mpc$^{-1}$. Left
panel shows the results derived from the PT template calculated in
$f(R)$ gravity. The three different contours represent the cases
with the PT template 
based on the TNS model [Eq.~(\ref{eq:TNS_model})]  
with and without $A$ and $B$ terms 
(magenta, green), and with $A$ and $B$
calculated in GR (blue), which are also shown in Fig.~\ref{fig:fR0_MCMC}.
On the other hand, in right panel, the results are shown
for the PT template calculated in GR. In GR, the
power spectrum template can be
written as functions of $k$, $\mu$, and the linear growth rate $f$, i.e.,
$P^{\rm(S)}(k,\mu;f)$.  Here, incorporating the linear growth rate of the
$f(R)$ gravity into the GR-based template, we derive the
constraints on $|f_{R,0}|$ and $\sigmav$, depicted as contour with orange color.
The contour with magenta color is the result
taking account of the
scale-dependent relative growth by introducing {\it gravity bias},
$\delta_{\rm n\mbox{-}body,F4}(\bfk)=b(k)\,\delta_{\rm PT,GR}(\bfk)$ with
$b(k)=(1+A_2\,k^2)/(1+A_1\,k)$ and marginalizing over the nuisance parameters
$A_1$ and $A_2$ [see Eq.~(\ref{eq:gravity_bias}) ].
\label{fig:2D_MCMC}}
\end{figure*}

\section{Implications}
\label{sec:implications}

Having confirmed that the PT model of RSD
works well at quasi-linear scales, in this section, we discuss
the potential impact of the PT template
on the RSD measurement at quasilinear scales.

In testing gravity with RSD, a primary goal would be to clarify whether
GR really holds on cosmological scales or not. In this respect, the
measurement of the linear growth rate, $f=d\ln D_+/d\ln a$,
provides an important clue, and with the
GR-based PT template, we may look for a possible deviation of $f$ from GR
prediction. One important property in a large class of modified gravity
models, including $f(R)$ gravity, is the scale dependence of $f$. Thus,
a detection of scale-dependent $f$ immediately implies the deviation of
gravity from GR. The crucial question is how well one can detect or
characterize such scale dependence in a model-independent manner.

Alternatively, we may consider some specific gravity models, and
try to directly constrain the models themselves. In this
case, the linear growth rate $f$ might not be an appropriate
indicator to characterize a possible deviation from GR. Rather, one
tries to directly constrain the model parameter of modified gravity
(e.g., $|f_{R,0}|$ in the case of $f(R)$ gravity). Then, with the prior
assumption of the specific gravity models, the question is
how well we can accurately constrain the model parameter based on the
PT template of RSD in an un-biased way.

Below, we will separately consider
these two issues, and examine the parameter estimation analysis. 
Note that 
we will adopt below $\chi^2$ of Eq.~(\ref{eq:chi2_red}) to estimate the 
goodness-of-fit. Strictly speaking, 
this is not entirely correct, because the nonlinear 
gravitational evolution induces the non-Gaussian contribution, which 
produces non-vanishing power spectrum covariances between 
diffrent Fourier modes. However, it is shown in the GR case that 
as long as we consider the quasi-linear scales at 
moderately high redshift, 
the off-diagonal componets are small enough, and the diagonal components 
can be approximately described by the simple Gaussian contribution, 
leading to a negligible influence on the parameter etstimation 
(e.g., \cite{Takahashi:2009bq,Takahashi:2009ty}). 
We thus expect that the same would be true in our case of 
the $f(R)$ gravity model that is close to GR, and 
our simple treatment with Gaussian error contribution 
would be validated at quasi-linear scales.

\subsection{Constraining model parameters of modified gravity}

Let us first consider the model-dependent analysis to constrain the model
parameter of modified gravity, assuming the
$f(R)$ gravity with $|f_{\rm R,0}|=10^{-4}$ as our fiducial gravity model.
For specific functional form with Eq.~(\ref{eq:model_approx})
[or Eq.~(\ref{eq:model})],
the parameter $|f_{\rm R,0}|$ is the only parameter characterizing a deviation
of gravity from GR. Thus, the test of gravity is made possible
with constraining the model parameter $|f_{\rm R,0}|$ by fitting the
theoretical template to the data set of redshift-space power spectrum.
Here, as a simple demonstration, we ignore the effect of galaxy bias, and
allowing $|f_{\rm R,0}|$ to flow, we
fit the PT template to the $N$-body data at $z=1$.

Fig.~\ref{fig:fR0_MCMC} shows the results of parameter estimation
based on the Markov chain Monte Carlo (MCMC) technique.
Assuming the hypothetical
survey limited by the cosmic variance error with the survey volume
$V=10\,h^{-3}\,$Gpc$^3$, the best-fit value of $|f_{\rm R,0}|$ and the
1-$\sigma$ statistical error are derived, and are plotted (top)
as function of maximum wavenumber, $k_{\rm max}$, together with the reduced
chi-squared statistic $\chi^2_{\rm red}$ (bottom), where $k_{\rm max}$ represents
the range of the wavenumber used for
parameter estimation. Note here that the number of free parameters is two,
i.e., $|f_{\rm R,0}|$ and $\sigmav$. Accordingly, the derived constraint
is rather tight, and a slight discrepancy between the template and data can
lead to a biased
estimation of the $|f_{\rm R,0}|$. Fig.~\ref{fig:fR0_MCMC} shows
that only the improved model of RSD computed in $f(R)$ gravity
(filled circles)
recovers the fiducial $|f_{\rm R,0}|$ out to $k_{\rm max}=0.15\,h$\,Mpc$^{-1}$,
corresponding to the applicable range of standard PT one-loop.
A slight change of the PT template, depicted as open circles and filled
triangles, leads to a biased estimation of the model parameter. Ignoring the
damping function $D_{\rm FoG}$ (crosses) further adds a large systematic
error. This is even true at $k_{\rm max}\lesssim0.1\,h$\,Mpc$^{-1}$.

Left panel of Fig.~\ref{fig:2D_MCMC} shows the representative result
of the two-dimensional constraints on $|f_{\rm R,0}|$ and $\sigmav$
taken from Fig.~\ref{fig:fR0_MCMC}, where we fix the maximum wavenumber to
$k_{\rm max}=0.15\,h$\,Mpc$^{-1}$.
The meaning of color types are the same as in Fig.~\ref{fig:fR0_MCMC}, and
in each error contour, inner and outer contours respectively
represent the $1$-$\sigma$ ($68\%$ C.L.) and $2$-$\sigma$ ($96\%$ C.L.)
constraints. Overall, the degeneracy between
$|f_{\rm R,0}|$ and $\sigmav$ is weak, and the result suggests that
at the scales accessible by PT template,
the model parameter $|f_{\rm R,0}|$ can be constrained down to
$\mathcal{O}(10^{-5})$ from future RSD measurements.

Note, however, that this is only true when
we properly take account of the effect of modified gravity in computing
the PT template. Most of the analysis in the literature considered the effect of
modified gravity only in the linear growth rate $f$ and incorporated
it into the GR-based template to constrain the model parameter $|f_{\rm R,0}|$
using the measurements of RSD (e.g.,
\cite{Yamamoto:2010ie,Okada:2012mn} for recent works).
The right panel of Fig.~\ref{fig:2D_MCMC} indeed demonstrates such a case. That is,
we adopt the GR-based PT template in which the effect of modified gravity
is only incorporated in the linear growth rate $f$. In GR,
the velocity-divergence field $\theta$ is known to be factorized as
$\theta(\bfk;t)=f\,\widetilde{\theta}(\bfk;t)$, where $\widetilde{\theta}$ is
perturbatively expanded as
$\widetilde{\theta}(\bfk;t)=\sum_n\,[D_+(t)]^n\,\widetilde{\theta}_n(\bfk)$.
As a result, at a given redshift, the PT template of the
redshift-space power spectrum is described as the function of $k$, $\mu$
and $f$, i.e., $P^{\rm(S)}(k,\mu;f)$. Since the growth rate $f$ controlls
the strength of RSD, we naively expect that simply
incorporating the scale-dependent $f$ in modified
gravity into the PT template allows us to faithfully constrain
the model parameter $|f_{\rm R,0}|$.

However, this actually leads to
a biased estimation of the model parameter $|f_{\rm R,0}|$,
as shown in the contour with orange color of Fig.~\ref{fig:2D_MCMC}.
The reason for the large systematic bias is
ascribed to the fact that the modification of gravity not only alters the
linear growth rate but also affects the shape of the real-space power spectra
because of the scale-dependent growth, as clearly shown in
Fig.~\ref{fig:pk_dd_dt_tt}. Thus, for an unbiased estimation of $|f_{\rm R,0}|$,
we need to additionally incorporate the effect of {\it gravity bias},
that accounts for the relative difference of the
clustering amplitude between GR and $f(R)$ gravity, into the PT template.
The contour with magenta color
is the results taking account of this gravity bias,
simply assuming the following relation:
\begin{align}
\delta_{\rm n\mbox{-}body,F4}(\bfk)=b(k)\,\delta_{\rm PT,GR}(\bfk);\quad
b(k) = \frac{1+A_2\,k^2}{1+A_1\,k},
\label{eq:gravity_bias}
\end{align}
where $\delta_{\rm n\mbox{-}body,F4}$ is the density field in $N$-body simulation,
whilst $\delta_{\rm GR}$ is the density field for the PT calculation. The
function $b(k)$ characterizes the scale-dependent growth relative to the
GR prediction, and we adopt here the functional form similar to those
frequently used to model the galaxy bias (e.g., \cite{Cole:2005sx,Sanchez:2007rc}).
Allowing the parameters $A_1$ and $A_2$ to float,
the result marginally reproduces the fiducial value of $|f_{\rm R,0}|$,
and the goodness-of-fit quantified by $\chi^2_{\rm red}$ is improved.
With the increased number of free parameters, however,
constraining power is significantly reduced, and  the
size of error contour indeed becomes large (c.f. left panel
of Fig.~\ref{fig:2D_MCMC}). This proves that the heterogeneous PT
template is insufficient to tightly constrain the model parameter
of modified gravity, and a full PT modeling taking proper account of
the modified gravity is required for unlocking the full power of
precision RSD measurement.

\begin{figure}[t]
\begin{center}
\hspace*{-0.5cm}
\includegraphics[width=9cm,angle=0]{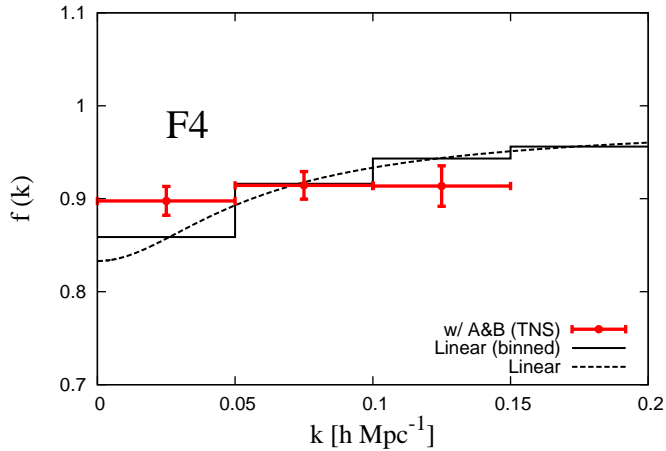}
\end{center}

\vspace*{-0.5cm}

\caption{MCMC results of the constraint on scale-dependent linear growth rate.
Using the GR-based PT template with the improved model of RSD,
we allow the linear growth rate $f$ to spatially vary in three wavenumber
bins. Adopting the gravity bias prescription given in
Eq.~(\ref{eq:gravity_bias}) and fixing the maximum wavenumber to
$k_{\rm max}=0.15\,h$\,Mpc$^{-1}$, we derive the constraint on $f$ in each
wavenumber bin. The vertical errorbars indicate the $1$-$\sigma$ error
assuming the cosmic-variance limited survey of $V=10\,h^{-3}\,$Gpc$^3$,
and the dotted and solid lines respectively 
represent the linear theory prediction and 
its binned average in the underlying $f(R)$ gravity model.
\label{fig:MCMC2}
}
\end{figure}

\begin{figure*}[ht]
\begin{center}
\hspace*{-0.5cm}
\includegraphics[width=9.5cm,angle=0]{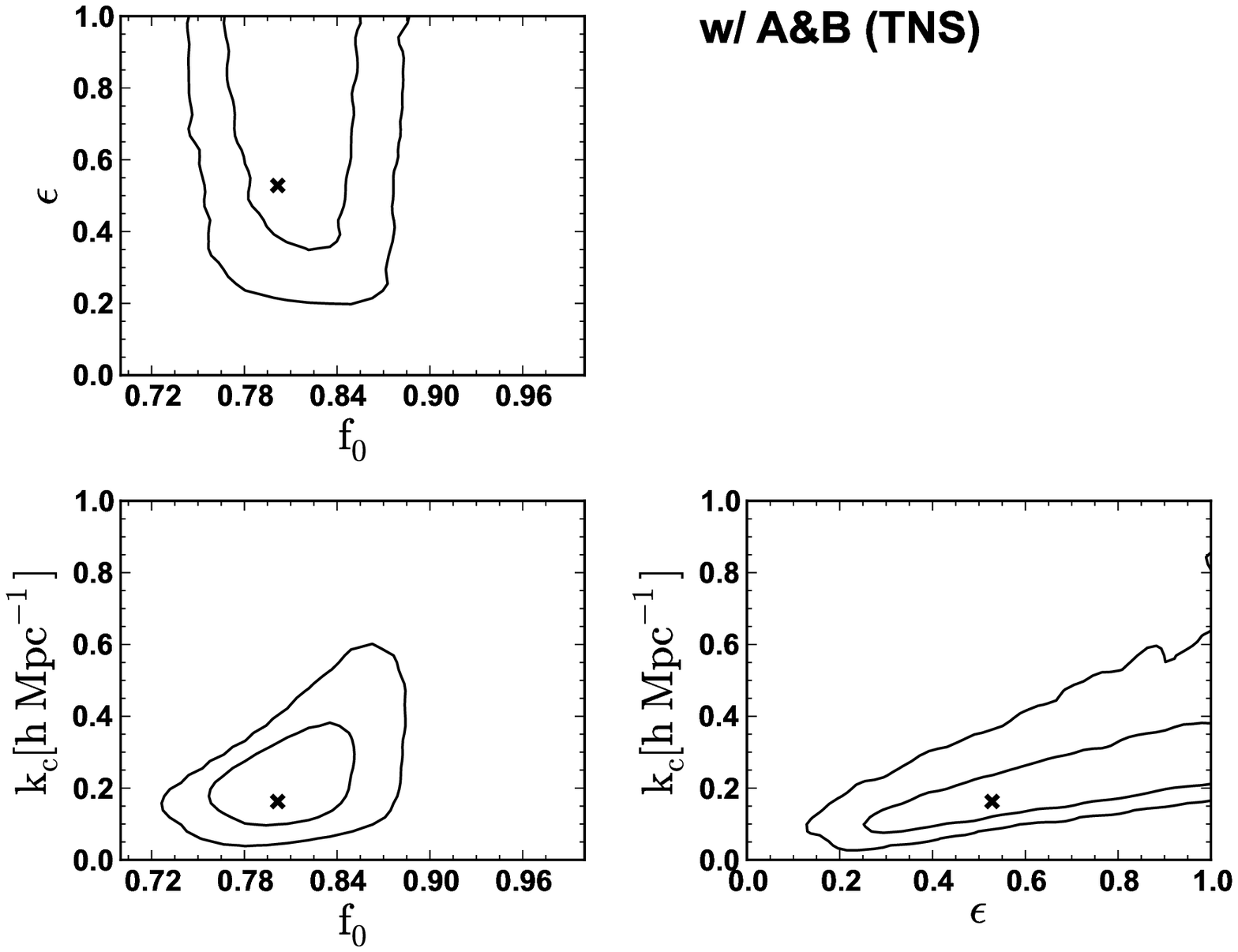}
\hspace*{-1.0cm}
\includegraphics[width=9.5cm,angle=0]{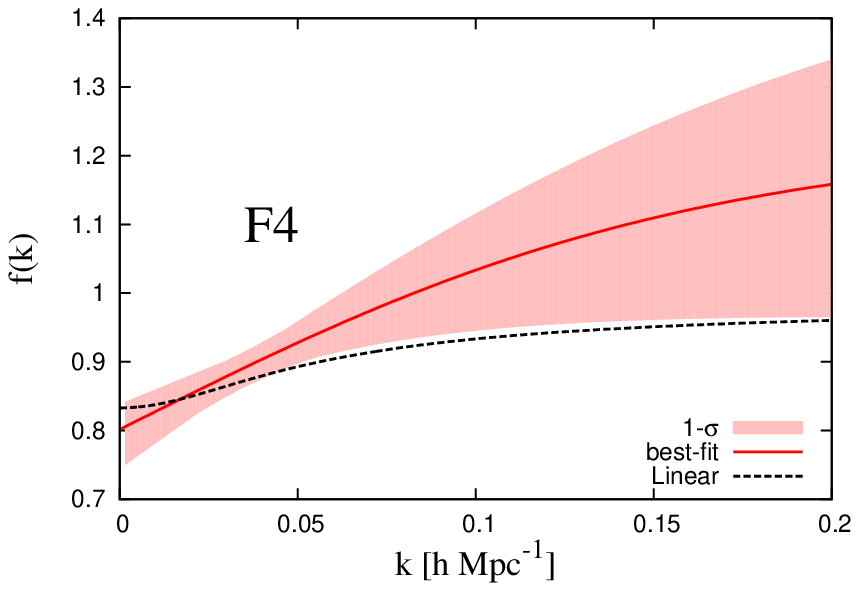}
\end{center}

\vspace*{-0.3cm}

\caption{MCMC results of the constraint on the scale-dependent linear growth rate,
assuming a specific functional form,
$f_{\rm approx}(k)=f_0[1+\epsilon\tanh(k/k_c)]$ [Eq.~(\ref{eq:f_approx})].
Using the GR-based PT template with the improved model of RSD, and adopting the
gravity bias in Eq.~(\ref{eq:gravity_bias}),
we derive the constraint on $f_0$ and $\epsilon$, and $k_c$. Left panel shows
the two-dimensional projected constraints, and the crosses indicate the
best-fit values. The inner and outer contours
respectively represent the $1$- and $2$-$\sigma$ statistical errors,
assuming the cosmic-variance limited survey of $V=10\,h^{-3}\,$Gpc$^3$.
In the right panel,
the best-fit curve of the scale-dependent linear growth is plotted
in a red solid line, and
its $1$-$\sigma$ statistical uncertainty is shown in
a red shaded region. For reference, the linear growth rate in the underlying
$f(R)$ gravity model is also plotted in a black dashed line.
\label{fig:MCMC2_fk_f0_float}
}
\end{figure*}

\subsection{Model-independent detection of a small deviation from GR}

Consider next the model-independent test of GR, and
discuss how well we can characterize or detect the scale dependence of
the linear growth rate, $f$. Here, for illustrative purpose, we examine the
two simple cases. One is to divide the power spectrum data
into several wavenumber bins, and in each bin, we try to estimate
$f$ to see a possible deviation from spatially homogeneous $f$.
The other case is to assume a specific functional form of $f$, and
to constrain its parameters. In both cases, similar to the
analysis shown in right panel of Fig.~\ref{fig:2D_MCMC},
we adopt the GR-based PT
template with an improved model of RSD (i.e., TNS model), 
and take account of the
gravity bias in Eq.~(\ref{eq:gravity_bias}). We then fit the template
to the monopole and quadrupole power spectra at $z=1$ measured from
$N$-body simulations of $f(R)$ gravity with $|f_{\rm R,0}|=10^{-4}$.

Fig.~\ref{fig:MCMC2} shows the result of MCMC analysis for
the binned linear growth rate,
where we set $k_{\rm max}=0.15\,h$\,Mpc$^{-1}$, and divide the power spectrum
data into three equal bins. Dotted and solid lines represents the 
linear growth rate of the $f(R)$ gravity with and without binning, 
while the vertical errorbars of the binned results
indicate the $1$-$\sigma$ statistical uncertainty derived from
the MCMC analysis, marginalized over other nuisance parameters.
Note that number of free parameters is $6$. The best-fit
value of $f$ in each bin is close to the fiducial value, but slightly away
from linear theory prediction except for the central bin. As a result,
the errorbars share almost the same value of $f$, and no notable trend
of the scale-dependent growth is found from the binned estimate of $f$.

Fig.~\ref{fig:MCMC2_fk_f0_float} examines the other case, in which
we assume a specific functional form of $f$ given below:
\begin{align}
f_{\rm approx}(k) = f_0\left[1+\epsilon\,\tanh(k/k_{\rm c})\right].
\label{eq:f_approx}
\end{align}
Allowing the parameters $f_0$, $\epsilon$ and $k_{\rm c}$ to float,
we perform the MCMC analysis. Again, the number of free parameters is $6$, 
and we set $k_{\rm max}=0.15\,h$\,Mpc$^{-1}$.
Note that for the scales of our interest, Eq.~(\ref{eq:f_approx}) is shown
to accurately describe the scale-dependent
linear growth rate of $f(R)$ gravity, and fitting directly
Eq.~(\ref{eq:f_approx}) to the linear theory prediction
at $z=1$, we obtain $f_0=0.83$,
$\epsilon=0.17$, and $k_c=0.11\,h$\,Mpc$^{-1}$.

The left panel of Fig.~\ref{fig:MCMC2_fk_f0_float}
shows the two-dimensional projected errors on the
parameters, $f_0$, $\epsilon$, and $k_c$.
The MCMC analysis of the RSD measurement
favors non-zero values of these parameters,
strongly indicating a deviation from spatially constant
$f$. However, a closer look at two-dimensional contours
reveals a substantial difference in $\epsilon$
between the best-fit result and the directly fitted value, $0.17$.
As a result, the MCMC result is unable to reproduce the
underlying scale-dependent linear growth rate. Right panel of
Fig.~\ref{fig:MCMC2_fk_f0_float} shows the constraint on the
scale-dependence of $f$. Based on the best-fit values and the associated
$1$-$\sigma$ errors shown in left panel of
Fig.~\ref{fig:MCMC2_fk_f0_float},
the best-fit curve is plotted in red solid line, and
its $1$-$\sigma$ statistical uncertainty is shown in
red shaded region. The scale-dependence inferred from the MCMC result
is rather stronger than that of the linear theory prediction (black dashed).

These two examples imply that the model-independent
detection and characterization of the scale-dependent $f$ are
generally difficult, and the results are rather sensitive to the
choice of parameterized form of the linear growth rate.
This is presumably because each of
the parameters characterizing the scale-dependent $f$ cannot be
determined locally, but rather it must be estimated with
a wide range of wavenumber. Then the parameters
tend to be highly correlated with each other,
leading to a biased estimation. In this respect, a sophisticated treatment
with principal component analysis may provide a way to
robustly detect a scale-dependent $f$.

\section{Conclusion}
\label{sec:conclusion}

In this paper, we studied how well we can clarify the nature of
gravity at large scales with redshift-space distortions (RSD), especially
focusing on the quasilinear regime of the gravitational evolution.
While most of previous works have been done with the theoretical template
assuming GR as underlying gravity theory, we here developed a new perturbation
theory (PT) prescription for RSD in the general context of the modified
gravity models. Extending our previous works on the improved model of RSD proposed by Ref.~\cite{Taruya:2010mx}, we applied the standard PT framework by
Ref.~\cite{Koyama:2009me}, which has been formulated to deal with a wide
class of modified gravity models, to the computation of the redshift-space power spectrum. As a specific application, in this paper, we consider the $f(R)$ gravity
model, and compared the PT prediction of RSD with results of $N$-body
simulations. Albeit the limited applicable range of the standard PT,
the PT results successfully describe the $N$-body simulations, and  the
predicted monopole and quadrupole spectra quantitatively agree with $N$-body results.

Then, we next considered how well we can characterize and/or constrain
the deviation of gravity from GR. One obvious approach is to first assume
a specific modified theory of gravity as an underlying gravity model and
constrain their model parameters. Using the PT as a theoretical template,
we performed the parameter estimation analysis, and checked if the theoretical
template correctly recovers the fiducial value of the model parameter in
the $N$-body simulations. Adopting the improved model of RSD 
[TNS model, Eq.~(\ref{eq:TNS_model})], a full PT
template calculated in the modified gravity model was found to reproduce the
correct model parameter, while a slight deficit in the PT template led to
a biased parameter estimation. As another approach, we have also examined
the model-independent analysis, and based on the PT template calculated
in GR, we tried to characterize the scale-dependent linear growth rate
from monopole and quadrupole power spectra. Without assuming any modified
gravity model, the parameterization of the scale-dependent linear growth rate
$f$ is necessary, and the parameters characterizing $f$
are highly correlated in general. Our simple two examples suggest that the
results are highly sensitive to the choice of parameterization, and
it is generally difficult to characterize the scale-dependence of $f$
in an unbiased manner unless employing some sophisticated methods such
as principal component analysis.

Throughout the paper, we have worked with the standard PT, but
the standard PT is known to have a bad convergence property. While
we can still get a fruitful constraint on modified gravity models,
resummed PT schemes with a wide applicable range are highly desirable to
improve the observational constraint. A development of improved PT template
in redshift space is an important future direction
(see \cite{Brax:2012sy,Brax:2013fna} for recent attempt).
Another important issue is the application of the present prescription to
the real measurement of RSD. With full PT implementation of the theoretical
template, a tight cosmological constraint is expected to be obtained in a robust
and unbiased way. In doing this, however, a proper account of
the galaxy bias would be crucial. Although the present paper mainly focused on
the matter power spectrum, the galaxy bias would be also affected by the
modification of gravity, and this may produce a non-trivial scale-dependent
shape of the observed power spectrum. 
In fact, the $N$-body study of 
the halo clustering properties has revealed that the halo bias 
in $f(R)$ gravity is systematically lower than that in GR 
(e.g., \cite{Schmidt:2008tn}). 
Since even the velocity dispersion and clustering amplitude of 
the dark matter distribution in $f(R)$ gravity differ from 
those in GR (see Figs.~\ref{fig:pk_dd_dt_tt} and \ref{fig:pkred} 
in Sec.~\ref{sec:RSD_fRmodel}), coupled with the nonlinear gravity, 
this could impose a non-trivial trend in the halo/galaxy bias 
(see e.g., \cite{Lombriser:2013wta, Ferraro:2010gh,Li:2011pj,Bianchi:2013foa} 
for recent study on the abundance and clustering of halos and galaxies). 
Hence, a careful study of the halo/galaxy bias is necessary 
 together with extensive tests with 
$N$-body mock catalogs toward an unbiased test of gravity.

\begin{acknowledgments}
We are grateful to Elise Jennings for providing us the power spectrum data of
$N$-body simulations. We also thank Baojiu Li and Gong-bo Zhao for allowing us 
to use the simulation data. This work is supported in part by a Grant-in-Aid for
Scientific Research from the Japan Society for the Promotion of Science 
(No.~23740186 for T.H and No.~24540257 for A.T). K.K is supported by STFC 
grant ST/K0090X/1, the European Research Council and the Leverhulme trust. 
T.H. acknowledges a support from MEXT HPCI Strategic Program. 
\end{acknowledgments}

\appendix
\section{Basic equations for perturbations and second-order kernels}
\label{sec:BasicEqs_MG}

In this appendix, after briefly reviewing the formalism developed in
Ref.~\cite{Koyama:2009me}, we derive general expressions for the second-order
PT kernel $F_a^{(2)}$, as well as the linear growth factor $F_a^{(1)}$ in
modified gravity models. In Sec.~\ref{subsec:evolution_eqs}, we begin
by reviewing the framework to treat the evolution of matter
fluctuations in modified gravity models. We then develop the perturbation
theory and derive the PT kernels up to the second-order in
Sec.~\ref{subsec:PT_kernel}. The explicit expressions for PT kernels are
given in specific modified gravity models, i.e., $f(R)$ gravity and
Dvali-Gabadadze-Porratti (DGP) models.

\subsection{Evolution equations}
\label{subsec:evolution_eqs}

Let us first consider the matter sector. Apart from the force law of gravity,
the basic equations governing the evolution of matter sector
is basically described by the conservation of energy momentum tensor,
which would remain unchanged even in the modified gravity model.
Hence, under the single-stream approximation, the matter fluctuations are
treated as pressureless fluid flow, whose evolution equations are the
continuity and Euler equations:
\begin{align}
&\frac{\partial \delta}{\partial t} +
\frac{1}{a}\nabla\cdot[(1+\delta)\bfv]=0,
\label{eq:eq_continuity}\\
&\frac{\partial \bfv}{\partial t} + H\,\bfv+
\frac{1}{a}(\bfv\cdot\nabla)\cdot\bfv=-\frac{1}{a}\nabla\psi,
\label{eq:eq_Euler}
\end{align}
where $\psi$ is the Newton potential.

On the other hand, for the gravity sector, there may appear
a new scalar degree of freedom referred to as the scalaron, which
results in a large-distance modification to the gravity. On large scales, the
scalaron $\varphi$ mediates the scalar force, and behaves like the Brans-Dicke
scalar field without potential and self-interactions, while it should
acquire some interaction terms on small scales, which play an
important role to recover GR. Indeed, for several known mechanisms
such as chameleon and Vainshtain mechanisms (e.g., \cite{Khoury:2003rn,Deffayet:2001uk}), the nonlinear
interaction terms naturally arise and eventually become dominant,
leading to a recovery of GR. As a result, even on subhorizon scales, the
Poisson equation is modified, and is coupled to the field equation for
scalaron $\varphi$ with self-interaction term:
\begin{align}
&\frac{1}{a}\nabla^2\psi=\frac{\kappa^2}{2}\,\rhom\,\delta
-\frac{1}{2a^2}\nabla^2\varphi,
\label{eq:Poisson_eq}\\
&(3+2\omega_{\rm BD})\frac{1}{a^2}\nabla^2\varphi=-2\kappa^2\rhom\,\delta
-\mathcal{I}(\varphi)
\label{eq:EoM_scalaron}
\end{align}
with $\kappa^2=8\pi\,G$ and $\omega_{\rm BD}$ being the Brans-Dicke parameter.
Here, we have used the quasi-static approximation and neglected the time
derivatives of the perturbed quantities compared with the spatial
derivatives. This treatment is always valid as long as we consider
the evolution of matter fluctuations inside the Hubble horizon.
The function $\mathcal{I}$ represents the nonlinear self-interaction,
and it can be expanded as
\begin{align}
&\mathcal{I}(\varphi)=M_1(k)+\frac{1}{2}\,
\int\frac{d^3\bfk_1 d^3\bfk_2}{(2\pi)^3}\delta_{\rm D}(\bfk-\bfk_{12})\,
\nonumber\\
&\qquad\qquad\times\,
M_2(\bfk_1,\bfk_2)\varphi(\bfk_1)\varphi(\bfk_2)+\cdots
\label{eq:I_expansion}
\end{align}

In Fourier space, Eqs.~(\ref{eq:eq_continuity})--(\ref{eq:EoM_scalaron})
can be reduced to a more compact form.
Assuming the irrotationality of fluid quantities,
the velocity field is expressed in terms of
velocity divergence, $\theta=\nabla\cdot\bfv/(aH)$.
Then, we have \cite{Koyama:2009me},
\begin{align}
&H^{-1}\frac{\partial \delta(\bfk)}{\partial t}+\theta(\bfk)
=-\int\frac{d^3\bfk_1d^3\bfk_2}{(2\pi)^3}\,\delta_D(\bfk-\bfk_{12})\,
\nonumber\\
&\qquad\qquad\qquad\qquad
\times\alpha(\bfk_1,\bfk_2)\,\theta(\bfk_1)\delta(\bfk_2),
\label{eq:EoM1}
\\
&H^{-1}\frac{\partial \theta(\bfk)}{\partial t}+
\left\{2+\frac{\dot{H}}{H^2}\right\}\theta(\bfk)+
\frac{\kappa^2\,\rho_{\rm m}}{2H^2}
\left\{1+\frac{1}{3}\frac{(k/a)^2}{\Pi(\bfk)}\right\}\delta(\bfk)
\nonumber\\
&\qquad
=-\frac{1}{2}\int\frac{d^3\bfk_1d^3\bfk_2}{(2\pi)^3}\,\delta_D(\bfk-\bfk_{12})\,
\beta(\bfk_1,\bfk_2)\,\theta(\bfk_1)\theta(\bfk_2)
\nonumber\\
&\quad\qquad-\frac{1}{2}\left(\frac{k}{a}\right)^2\,\frac{S(\bfk)}{H^2},
\label{eq:EoM2}
\end{align}
with the mode-coupling kernels, $\alpha$ and $\beta$ given by
\begin{align}
&\alpha(\bfk_1,\bfk_2)=1+\frac{\bfk_1\cdot\bfk_2}{|\bfk_1|^2},
\nonumber\\
&\beta(\bfk_1,\bfk_2)=\frac{(\bfk_1\cdot\bfk_2)|\bfk_1+\bfk_2|^2}
{|\bfk_1|^2|\bfk_2|^2}.
\nonumber
\end{align}
In the above, the function $\Pi$ characterizes the
deviation of the Newton constant from GR,
while the quantity $S$ is originated from
the non-linear interactions of the scalaron, which is
responsible for the recovery of GR at small scales.
The functional form of these
are obtained from the Poisson equation and field equation for scalaron,
and the expressions relevant for the second-order perturbations are
respectively given by \cite{Koyama:2009me}:
\begin{align}
&\Pi(k)=\frac{1}{3}\left\{(3+2\omega_{\rm BD})\frac{k^2}{a^2}+M_1(k)\right\},
\nonumber\\
&S(k)=-\frac{1}{6\,\Pi(k)} \left(\frac{\kappa^2\,\rho_{\rm m}}{3}\right)^2
\int\frac{d^3\bfk_1d^3\bfk_2}{(2\pi)^3}\,\delta_D(\bfk-\bfk_{12})\,
\nonumber\\
&\qquad\times
M_2(\bfk_1,\bfk_2)\frac{\delta(\bfk_1)\delta(\bfk_2)}{\Pi(k_1)\Pi(k_2)}
 + \cdots.
\end{align}
Here, in deriving the last expression,
we perturbatively solve the scalaron field $\varphi$ in terms of $\delta$
using Eqs.~(\ref{eq:EoM_scalaron}) and (\ref{eq:I_expansion}).
The explicit functional form of $\Pi$ or $M_1$, and $M_2$ depends
on actual modified gravity models, which will be specified later.

\subsection{PT kernels}
\label{subsec:PT_kernel}

Let us now perturbatively solve the evolution equations (\ref{eq:EoM1}) and
(\ref{eq:EoM2}).
Consider first the linear-order solutions. Ignoring all the non-linear
terms in Eqs.~(\ref{eq:EoM1}) and (\ref{eq:EoM2}), we obtain
\begin{align}
&\ddot{\delta}^{(1)}(\bfk)+2H\,\dot{\delta}^{(1)}(\bfk)-
\frac{\kappa^2\,\rho_{\rm m}}{2}
\left\{1+\frac{1}{3}\frac{(k/a)^2}{\Pi(k)}\right\}\delta^{(1)}(\bfk)=0,
\nonumber\\
&\theta^{(1)}(\bfk)=-\frac{1}{H}\dot{\delta}^{(1)}(\bfk)
\label{eq:EoM_lin}
\end{align}
The solutions $\delta^{(1)}$ and $\theta^{(1)}$ can be formally expressed as
\begin{align}
\delta^{(1)}(\bfk;t)=D(k;t)\,\delta_0(\bfk),
\quad
\theta^{(1)}(\bfk;t)=-\frac{\dot{D}(k;t)}{H}\,\delta_0(\bfk),
\label{eq:linear_solutions}
\end{align}
where the function $\delta_0(\bfk)$ is the initial density field
[see Eq.~(\ref{eq:initial_pk})].
The function $D$ is the linear growth factor, and satisfies
the following evolution equation:
\begin{align}
\ddot{D}+2H\,\dot{D}-
\frac{\kappa^2\,\rho_{\rm m}}{2}
\left\{1+\frac{1}{3}\frac{(k/a)^2}{\Pi(k)}\right\}D=0.
\end{align}
Accordingly, the first-order PT kernels are
\begin{align}
F_1^{(1)}(k;t)=D(k;t), \quad F_2^{(1)}(k;t)=-\frac{\dot{D}(k;t)}{H}.
\end{align}

Next consider the second-order solutions.
Substituting the linear-order solutions into the right-hand side of
Eqs.~(\ref{eq:EoM1}) and (\ref{eq:EoM2}), the equations for second-order
perturbations are
\begin{widetext}
\begin{align}
&\ddot{\delta}^{(2)}(\bfk)+2H\,
\dot{\delta}^{(2)}(\bfk)-\frac{\kappa^2\,\rho_{\rm m}}{2}
\left\{1+\frac{1}{3}\frac{(k/a)^2}{\Pi(k)}\right\}\delta^{(2)}(\bfk)
=\,\,
\int\frac{d^3\bfk_1d^3\bfk_2}{(2\pi)^3}\,\delta_D(\bfk-\bfk_{12})
\nonumber
\\
&\quad\times
\Bigl[\left\{(\ddot{D}_1+2H\dot{D}_1)\,\,D_2+
\dot{D}_1\dot{D}_2\do\right\}\alpha_{1,2}+
\frac{\dot{D}_1\dot{D}_2}{2}\beta_{1,2}-
\frac{(k_{12}/a)^2}{12\Pi(k_{12})}
\left(\frac{\kappa^2\rho_{\rm m}}{3}\right)^2
\frac{M_2(\bfk_1,\bfk_2)}{\Pi_1\,\Pi_2}\,D_1\,D_2
\Bigr]\,\delta_0(\bfk_1)\delta_0(\bfk_2),
\label{eq:EoM_2nd_PT}\\
&\theta^{(2)}(\bfk)=-\frac{1}{H}\dot{\delta}^{(2)}(\bfk)+\frac{1}{H}
\int\frac{d^3\bfk_1d^3\bfk_2}{(2\pi)^3}\,\delta_D(\bfk-\bfk_{12})\,
\alpha(\bfk_1,\bfk_2)\,\dot{D}_1\,D_2 \,\delta_0(\bfk_1)\,\delta_0(\bfk_2)
\end{align}
\end{widetext}
where we introduced the short-hand notations, $D_i=D(k_i;t)$,
$\alpha_{1,2}=\alpha(\bfk_1,\bfk_2)$, and $\Pi_i=\Pi(k_i)$.
Then, the second-order PT solutions are formally written as
\begin{widetext}
\begin{align}
&\delta^{(2)}(\bfk;t) = \int\frac{d^3\bfk_1d^32\bfk_2}{(2\pi)^3}
\delta_D(\bfk-\bfk_{12})\,\,
\left[\frac{1}{2}\left(D_{1,2}^{(2)}\,\alpha_{1,2}+D_{2,1}^{(2)}\,\alpha_{2,1}\right)+
E^{(2)}_{1,2}\,\beta_{1,2}+F^{(2)}_{1,2}\right]\,\delta_0(\bfk_1)\,\delta_0(\bfk_2),
\label{eq:2nd_order_solutions1}
\\
&\theta^{(2)}(\bfk;t) = \frac{1}{H}\int\frac{d^3\bfk_1d^32\bfk_2}{(2\pi)^3}
\delta_D(\bfk-\bfk_{12})
\nonumber\\
&\qquad\qquad\qquad\times
\left[\frac{1}{2}\left\{\left(\dot{D}_1D_2-\dot{D}^{(2)}_{1,2}\right)\alpha_{1,2}+
\left(\dot{D}_2D_1-\dot{D}^{(2)}_{2,1}\right)\alpha_{2,1}\right\}
-\dot{E}^{(2)}_{1,2}\,\beta_{1,2}-\dot{F}^{(2)}_{1,2}\right]\,
\delta_0(\bfk_1)\,\delta_0(\bfk_2).
\label{eq:2nd_order_solutions2}
\end{align}
\end{widetext}
Thus, the {\it symmetrized} second-order PT kernels $F_a^{(2)}$ are respectively given by
\begin{align}
&F_1^{(2)}(\bfk_1,\bfk_2;t)=
\frac{1}{2}\left(D_{1,2}^{(2)}\,\alpha_{1,2}+D_{2,1}^{(2)}\,\alpha_{2,1}\right)
\nonumber\\
&\qquad\qquad+E^{(2)}_{1,2}\,\beta_{1,2}+F^{(2)}_{1,2},
\label{eq:F2_density}
\\
&F_2^{(2)}(\bfk_1,\bfk_2;t)=\frac{1}{H}\Bigl[\,
\frac{1}{2}\Bigl\{\left(\dot{D}_1D_2-\dot{D}^{(2)}_{1,2}\right)\alpha_{1,2}
\nonumber
\\
&+\left(\dot{D}_2D_1-\dot{D}^{(2)}_{2,1}\right)\alpha_{2,1}\Bigr\}
-\dot{E}^{(2)}_{1,2}\,\beta_{1,2}-\dot{F}^{(2)}_{1,2}
\,\Bigr].
\label{eq:F2_velocity}
\end{align}
Here, the functions $D^{(2)}_{1,2}=D^{(2)}(\bfk_1,\bfk_2)$,
$E^{(2)}_{1,2}=E^{(2)}(\bfk_1,\bfk_2)$, and
$F^{(2)}_{1,2}=F^{(2)}(\bfk_1,\bfk_2)$ are the second-order growth functions,
whose governing equations are
\begin{align}
&\hat{\mathcal{L}}\,\,D^{(2)}(\bfk_1,\bfk_2)
\nonumber\\
&\,\,=
\left\{\ddot{D}(k_1)+2H\dot{D}(k_1)\right\}\,D(k_2)
+\dot{D}(k_1)\dot{D}(k_2),
\label{eq:EoM_D2}
\\
&\hat{\mathcal{L}}\,\,E^{(2)}(\bfk_1,\bfk_2)=
\frac{1}{2}\dot{D}(k_1)\dot{D}(k_2),
\label{eq:EoM_E2}
\\
&\hat{\mathcal{L}}\,\,F^{(2)}(\bfk_1,\bfk_2)
\nonumber\\
&=-\frac{(k_{12}/a)^2}{12\,\Pi(k_{12})}
\left(\frac{\kappa^2\,\rho_{\rm m}}{3}\right)^2
\frac{M_2(\bfk_1,\bfk_2)}{\Pi(k_1)\Pi(k_2)}\,D(k_1)D(k_2)
\label{eq:EoM_F2}
\end{align}
with the operator $\hat{\mathcal{L}}$ given by
\begin{align}
&\hat{\mathcal{L}}(k_{12},t)
\equiv\frac{d^2}{dt^2}+2H\frac{d}{dt}-
\frac{\kappa^2\,\rho_{\rm m}}{2}
\left\{1+\frac{1}{3}\frac{(k_{12}/a)^2}{\Pi(k_{12})}\right\}.
\end{align}

Below, we will present a more explicit expression for evolution equations of
the growth functions in $f(R)$ gravity and DGP models.

\subsubsection{$f(R)$ gravity models}
\label{subsubsec:f(R)}

In $f(R)$ gravity models described in Eq.~(\ref{eq:action_fR}),
the scalaron $\varphi$ is identified with $\varphi=f_R-\overline{f}_R$
[Eq.~(\ref{eq:scalaron_fR})], and it behaves like the Brans-Dicke scalar with
$\omega_{\rm BD}=0$ and the nonlinear interaction
$\mathcal{I}(\varphi)=R(f_R)-R(\overline{f}_R)$. Then,
the functions $\Pi$ and $M_2$ are generally given by \cite{Koyama:2009me}
\begin{align}
&\Pi(k)=\left(\frac{k}{a}\right)^2+\frac{\overline{R}_{,f}(t)}{3},
\nonumber\\
&M_2(\bfk_1,\bfk_2) = \overline{R}_{,ff},
\nonumber
\end{align}
where we define $\overline{R}_{,f}=d\overline{R}(f_R)/df_R$ and
$\overline{R}_{,ff}=d^2\overline{R}(f_R)/df_R^2$. In $f(R)$ gravity,
all the second-order growth functions $D^{(2)}$, $E^{(2)}$, and $F^{(2)}$
as well as the linear growth factor $D$ are scale-dependent, and
no simplified expressions are obtained without invoking any approximations.
We numerically solve evolution equations below:
\begin{widetext}
\begin{align}
&  \hat{\mathcal{L}}_f(k,t)D(k)=0,
\label{eq:EoM_D_fR}
\\
&  \hat{\mathcal{L}}_f(k_{12},t)D^{(2)}(\bfk_1,\bfk_2)=
\dot{D}(k_1)\dot{D}(k_2)
+\frac{\kappa^2\,\rho_{\rm m}}{2}
\left\{1+\frac{1}{3}\frac{(k_1/a)^2}{\overline{R}_{,f}/3+(k_1/a)^2}\right\}
D(k_1)D(k_2),
\label{eq:EoM_D2_fR}
\\
&  \hat{\mathcal{L}}_f(k_{12},t)E^{(2)}(\bfk_1,\bfk_2)=
\frac{1}{2}\dot{D}(k_1)\dot{D}(k_2),
\label{eq:EoM_E2_fR}
\\
&  \hat{\mathcal{L}}_f(k_{12},t)F^{(2)}(\bfk_1,\bfk_2)=
-\frac{1}{12}\left(\frac{\kappa^2\,\rho_{\rm m}}{3}\right)^2
\frac{(k_{12}/a)^2}{\overline{R}_{,f}/3+(k_{12}/a)^2}
\frac{\overline{R}_{,ff}}{\{\overline{R}_{,f}/3+(k_1/a)^2\}\{\overline{R}_{,f}/3+(k_2/a)^2\}}\,D(k_1)D(k_2)
\label{eq:EoM_F2_fR}
\end{align}
\end{widetext}
with the linear operator being
\begin{align}
&\hat{\mathcal{L}}_f(k,t)\equiv\frac{d^2}{dt^2}+2H\frac{d}{dt}-
\frac{\kappa^2\,\rho_{\rm m}}{2}
\left\{1+\frac{1}{3}\frac{(k/a)^2}{\overline{R}_{,f}/3+(k/a)^2}\right\}.
\end{align}
Here, the function $D^{(2)}$ is asymmetric with respect to the change
of the arguments, i.e., $D^{(2)}(\bfk_1,\bfk_2)\neq D^{(2)}(\bfk_2,\bfk_1)$.
The second-order growth functions are generally given as the
function of $k_1$, $k_2$, and $k_{12}=|k_1^2+k_2^2+2(\bfk_1\cdot\bfk_2)|^{1/2}$.
Note that in deriving the evolution equations above,
we did not specify the functional form of $f(R)$, and
the functions, $\overline{R}_{,f}$ and $\overline{R}_{,ff}$, still remains
unspecified. To solve the equations, in this paper, we consider the specific
function given in Eq.~(\ref{eq:model}), and then the functions
$\overline{R}_{,f}$ and $\overline{R}_{,ff}$ are
expressed in terms of the background quantities.

\subsubsection{DGP model}
\label{subsec:DGP}

As another example of modified gravity model,
we consider the Dvali-Gabadadze-Porratti (DGP) braneworld model
\cite{Dvali:2000hr}. In DGP models, the Brans-Dicke parameter of the
scalaron becomes time-dependent, and is given by \cite{Koyama:2009me}
\begin{align}
&\omega_{\rm BD}=\frac{3}{2}(\beta-1),
\nonumber\\
&\beta=1-2\epsilon\,H\,r_c\left(1+\frac{\dot{H}}{3H^2}\right),
\end{align}
with $\epsilon=\pm1$, which represents two distinct branches of the
background solutions ($\epsilon=+1$: self-accelerating branch,
$\epsilon=-1:$ normal branch). The parameter $r_c$ is the crossover
scale which characterizes the ratio of $5D$ Newton constant to $4D$
Newton constant. In this model, the nonlinear interaction of the scalaron
comes from the Vainshtein mechanism. As a result,
the functions $\Pi$ and $M_2$ are respectively given by \cite{Koyama:2009me}
\begin{align}
&\Pi(\bfk)=\beta(t)\,\left(\frac{k}{a}\right)^2,
\nonumber\\
&M_2(\bfk_1,\bfk_2) = 2\frac{r_c^2}{a^4}
\left\{
k_1^2\,k_2^2-(\bfk_1\cdot\bfk_2)^2.
\right\},
\nonumber
\end{align}
Then, the second-order growth functions $D^{(2)}$, $E^{(2)}$, and $F^{(2)}$
as well as the linear growth factor $D$ become all independent of scale.

To further get a simplified expression, we may employ the
Einstein-de Sitter approximation.  In this approximation, the
non-linear growth functions in the higher-order PT solutions are first
obtained assuming the Einstein-de Sitter background, and they
are expressed in terms of the scale factor. Then, simply replacing the
scale factor with the linear growth factor $D(t)$, we obtain an approximate
description of the non-linear growth functions:
\begin{align}
D^{(2)}\to\frac{5}{7}\,D^2(t), \qquad
E^{(2)}\to\frac{1}{7}\,D^2(t).
\label{eq:D2_E2_EdS}
\end{align}
Here, the evolution equation for linear growth factor is given by
\begin{align}
&\ddot{D}+2H\dot{D}-
\frac{\kappa^2\,\rho_{\rm m}}{2}\left(1+\frac{1}{3\beta}\right)D=0.
\end{align}
Substituting Eq.~(\ref{eq:D2_E2_EdS}) into Eqs.~(\ref{eq:F2_density}) and
(\ref{eq:F2_velocity}), we obtain the approximate expressions for
the symmetrized PT kernels:
\begin{align}
&F_1^{(2)}(\bfk_1,\bfk_2;t)
\nonumber\\
&=D^2(t)\left\{
\frac{5}{14}\,(\alpha_{1,2}+\alpha_{2,1})+\frac{1}{7}\,\beta_{1,2}
\right\}+\left(1-\mu_{1,2}^2\right)\,\widetilde{F}_2(t),
\label{eq:F2_density_DGP}
\\
&F_2^{(2)}(\bfk_1,\bfk_2;t)=-\frac{D(t)\dot{D}(t)}{H}
\nonumber\\
&\quad\times
\left\{
\frac{3}{14}\,(\alpha_{1,2}+\alpha_{2,1})+\frac{2}{7}\,\beta_{1,2}
\right\}-\left(1-\mu_{1,2}^2\right)\,\frac{\dot{f}^{(2)}(t)}{H},
\label{eq:F2_velocity_DGP}
\end{align}
with $\mu_{1,2}=(\bfk_1\cdot\bfk_2)/(k_1k_2)$. In the above, we rewrite the
second-order growth function $F^{(2)}$ as
$F^{(2)}(\bfk_1,\bfk_2)=(1-\mu_{1,2}^2)\,f^{(2)}$, with
$f^{(2)}$ being the scale-independent function satisfying
the following evolution equation:
\begin{align}
&\ddot{f}^{(2)}+2H\dot{f}^{(2)}-
\frac{\kappa^2\,\rho_{\rm m}}{2}\left(1+\frac{1}{3\beta}\right)f^{(2)}
\nonumber\\
&\qquad\qquad\qquad\qquad=
-\frac{r_c^2}{6\beta^3}\left(\frac{\kappa^2\,\rho_{\rm m}}{3}\right)^2\,D^2(t).
\end{align}

Eqs.~(\ref{eq:F2_density_DGP}) and (\ref{eq:F2_velocity_DGP}) coincide with
the results in (B5) and (B6) of Ref.~\cite{Koyama:2009me}. The first term
of the right-hand-side in each kernel is exactly the same kernel as found
in GR, while the second term is originated from
the non-linear interactions of the scalar degree of freedom.


\begin{thebibliography}{79}
\expandafter\ifx\csname natexlab\endcsname\relax\def\natexlab#1{#1}\fi
\expandafter\ifx\csname bibnamefont\endcsname\relax
  \def\bibnamefont#1{#1}\fi
\expandafter\ifx\csname bibfnamefont\endcsname\relax
  \def\bibfnamefont#1{#1}\fi
\expandafter\ifx\csname citenamefont\endcsname\relax
  \def\citenamefont#1{#1}\fi
\expandafter\ifx\csname url\endcsname\relax
  \def\url#1{\texttt{#1}}\fi
\expandafter\ifx\csname urlprefix\endcsname\relax\def\urlprefix{URL }\fi
\providecommand{\bibinfo}[2]{#2}
\providecommand{\eprint}[2][]{\url{#2}}

\bibitem[{\citenamefont{Hamilton}(1997)}]{Hamilton:1997zq}
\bibinfo{author}{\bibfnamefont{A.~J.~S.} \bibnamefont{Hamilton}}
  (\bibinfo{year}{1997}), \eprint{astro-ph/9708102}.

\bibitem[{\citenamefont{Peebles}(Princeton University Press,
  1980)}]{Peebles:1980}
\bibinfo{author}{\bibfnamefont{P.}~\bibnamefont{Peebles}},
  \bibinfo{journal}{{\it The large-scale structure of the universe}}
  (\bibinfo{year}{Princeton University Press, 1980}).

\bibitem[{\citenamefont{Seo and Eisenstein}(2003)}]{Seo:2003pu}
\bibinfo{author}{\bibfnamefont{H.-J.} \bibnamefont{Seo}} \bibnamefont{and}
  \bibinfo{author}{\bibfnamefont{D.~J.} \bibnamefont{Eisenstein}},
  \bibinfo{journal}{Astrophys. J.} \textbf{\bibinfo{volume}{598}},
  \bibinfo{pages}{720} (\bibinfo{year}{2003}), \eprint{astro-ph/0307460}.

\bibitem[{\citenamefont{Blake and Glazebrook}(2003)}]{Blake:2003rh}
\bibinfo{author}{\bibfnamefont{C.}~\bibnamefont{Blake}} \bibnamefont{and}
  \bibinfo{author}{\bibfnamefont{K.}~\bibnamefont{Glazebrook}},
  \bibinfo{journal}{Astrophys. J.} \textbf{\bibinfo{volume}{594}},
  \bibinfo{pages}{665} (\bibinfo{year}{2003}), \eprint{astro-ph/0301632}.

\bibitem[{\citenamefont{Matsubara}(2004)}]{Matsubara:2004fr}
\bibinfo{author}{\bibfnamefont{T.}~\bibnamefont{Matsubara}},
  \bibinfo{journal}{Astrophys.J.} \textbf{\bibinfo{volume}{615}},
  \bibinfo{pages}{573} (\bibinfo{year}{2004}), \eprint{astro-ph/0408349}.

\bibitem[{\citenamefont{Glazebrook and Blake}(2005)}]{Glazebrook:2005mb}
\bibinfo{author}{\bibfnamefont{K.}~\bibnamefont{Glazebrook}} \bibnamefont{and}
  \bibinfo{author}{\bibfnamefont{C.}~\bibnamefont{Blake}},
  \bibinfo{journal}{Astrophys. J.} \textbf{\bibinfo{volume}{631}},
  \bibinfo{pages}{1} (\bibinfo{year}{2005}), \eprint{astro-ph/0505608}.

\bibitem[{\citenamefont{Linder}(2008)}]{Linder:2007nu}
\bibinfo{author}{\bibfnamefont{E.~V.} \bibnamefont{Linder}},
  \bibinfo{journal}{Astropart. Phys.} \textbf{\bibinfo{volume}{29}},
  \bibinfo{pages}{336} (\bibinfo{year}{2008}), \eprint{0709.1113}.

\bibitem[{\citenamefont{Guzzo et~al.}(2008)}]{Guzzo:2008ac}
\bibinfo{author}{\bibfnamefont{L.}~\bibnamefont{Guzzo}} \bibnamefont{et~al.},
  \bibinfo{journal}{Nature} \textbf{\bibinfo{volume}{451}},
  \bibinfo{pages}{541} (\bibinfo{year}{2008}), \eprint{0802.1944}.

\bibitem[{\citenamefont{Yamamoto et~al.}(2008)\citenamefont{Yamamoto, Sato, and
  Huetsi}}]{Yamamoto:2008gr}
\bibinfo{author}{\bibfnamefont{K.}~\bibnamefont{Yamamoto}},
  \bibinfo{author}{\bibfnamefont{T.}~\bibnamefont{Sato}}, \bibnamefont{and}
  \bibinfo{author}{\bibfnamefont{G.}~\bibnamefont{Huetsi}},
  \bibinfo{journal}{Prog. Theor. Phys.} \textbf{\bibinfo{volume}{120}},
  \bibinfo{pages}{609} (\bibinfo{year}{2008}), \eprint{0805.4789}.

\bibitem[{\citenamefont{Song and Percival}(2009)}]{Song:2008qt}
\bibinfo{author}{\bibfnamefont{Y.-S.} \bibnamefont{Song}} \bibnamefont{and}
  \bibinfo{author}{\bibfnamefont{W.~J.} \bibnamefont{Percival}},
  \bibinfo{journal}{JCAP} \textbf{\bibinfo{volume}{0910}}, \bibinfo{pages}{004}
  (\bibinfo{year}{2009}), \eprint{0807.0810}.

\bibitem[{\citenamefont{Song and Kayo}(2010)}]{Song:2010bk}
\bibinfo{author}{\bibfnamefont{Y.-S.} \bibnamefont{Song}} \bibnamefont{and}
  \bibinfo{author}{\bibfnamefont{I.}~\bibnamefont{Kayo}}
  (\bibinfo{year}{2010}), \eprint{1003.2420}.

\bibitem[{\citenamefont{Guzik et~al.}(2010)\citenamefont{Guzik, Jain, and
  Takada}}]{Guzik:2009cm}
\bibinfo{author}{\bibfnamefont{J.}~\bibnamefont{Guzik}},
  \bibinfo{author}{\bibfnamefont{B.}~\bibnamefont{Jain}}, \bibnamefont{and}
  \bibinfo{author}{\bibfnamefont{M.}~\bibnamefont{Takada}},
  \bibinfo{journal}{Phys.Rev.} \textbf{\bibinfo{volume}{D81}},
  \bibinfo{pages}{023503} (\bibinfo{year}{2010}), \eprint{0906.2221}.

\bibitem[{\citenamefont{Song et~al.}(2011)\citenamefont{Song, Zhao, Bacon,
  Koyama, Nichol et~al.}}]{Song:2010fg}
\bibinfo{author}{\bibfnamefont{Y.-S.} \bibnamefont{Song}},
  \bibinfo{author}{\bibfnamefont{G.-B.} \bibnamefont{Zhao}},
  \bibinfo{author}{\bibfnamefont{D.}~\bibnamefont{Bacon}},
  \bibinfo{author}{\bibfnamefont{K.}~\bibnamefont{Koyama}},
  \bibinfo{author}{\bibfnamefont{R.~C.} \bibnamefont{Nichol}},
  \bibnamefont{et~al.}, \bibinfo{journal}{Phys.Rev.}
  \textbf{\bibinfo{volume}{D84}}, \bibinfo{pages}{083523}
  (\bibinfo{year}{2011}), \eprint{1011.2106}.

\bibitem[{\citenamefont{Asaba et~al.}(2013)\citenamefont{Asaba, Hikage, Koyama,
  Zhao, Hojjati et~al.}}]{Asaba:2013xql}
\bibinfo{author}{\bibfnamefont{S.}~\bibnamefont{Asaba}},
  \bibinfo{author}{\bibfnamefont{C.}~\bibnamefont{Hikage}},
  \bibinfo{author}{\bibfnamefont{K.}~\bibnamefont{Koyama}},
  \bibinfo{author}{\bibfnamefont{G.-B.} \bibnamefont{Zhao}},
  \bibinfo{author}{\bibfnamefont{A.}~\bibnamefont{Hojjati}},
  \bibnamefont{et~al.}, \bibinfo{journal}{JCAP}
  \textbf{\bibinfo{volume}{1308}}, \bibinfo{pages}{029} (\bibinfo{year}{2013}),
  \eprint{1306.2546}.

\bibitem[{\citenamefont{Perlmutter et~al.}(1999)}]{Perlmutter:1998np}
\bibinfo{author}{\bibfnamefont{S.}~\bibnamefont{Perlmutter}}
  \bibnamefont{et~al.} (\bibinfo{collaboration}{Supernova Cosmology Project}),
  \bibinfo{journal}{Astrophys. J.} \textbf{\bibinfo{volume}{517}},
  \bibinfo{pages}{565} (\bibinfo{year}{1999}), \eprint{astro-ph/9812133}.

\bibitem[{\citenamefont{Riess et~al.}(1998)}]{Riess:1998cb}
\bibinfo{author}{\bibfnamefont{A.~G.} \bibnamefont{Riess}} \bibnamefont{et~al.}
  (\bibinfo{collaboration}{Supernova Search Team}), \bibinfo{journal}{Astron.
  J.} \textbf{\bibinfo{volume}{116}}, \bibinfo{pages}{1009}
  (\bibinfo{year}{1998}), \eprint{astro-ph/9805201}.

\bibitem[{\citenamefont{Khoury and Weltman}(2004)}]{Khoury:2003rn}
\bibinfo{author}{\bibfnamefont{J.}~\bibnamefont{Khoury}} \bibnamefont{and}
  \bibinfo{author}{\bibfnamefont{A.}~\bibnamefont{Weltman}},
  \bibinfo{journal}{Phys.Rev.} \textbf{\bibinfo{volume}{D69}},
  \bibinfo{pages}{044026} (\bibinfo{year}{2004}), \eprint{astro-ph/0309411}.

\bibitem[{\citenamefont{Deffayet et~al.}(2002)\citenamefont{Deffayet, Dvali,
  Gabadadze, and Vainshtein}}]{Deffayet:2001uk}
\bibinfo{author}{\bibfnamefont{C.}~\bibnamefont{Deffayet}},
  \bibinfo{author}{\bibfnamefont{G.}~\bibnamefont{Dvali}},
  \bibinfo{author}{\bibfnamefont{G.}~\bibnamefont{Gabadadze}},
  \bibnamefont{and} \bibinfo{author}{\bibfnamefont{A.~I.}
  \bibnamefont{Vainshtein}}, \bibinfo{journal}{Phys.Rev.}
  \textbf{\bibinfo{volume}{D65}}, \bibinfo{pages}{044026}
  (\bibinfo{year}{2002}), \eprint{hep-th/0106001}.

\bibitem[{\citenamefont{Hu and Sawicki}(2007)}]{Hu:2007nk}
\bibinfo{author}{\bibfnamefont{W.}~\bibnamefont{Hu}} \bibnamefont{and}
  \bibinfo{author}{\bibfnamefont{I.}~\bibnamefont{Sawicki}},
  \bibinfo{journal}{Phys.Rev.} \textbf{\bibinfo{volume}{D76}},
  \bibinfo{pages}{064004} (\bibinfo{year}{2007}), \eprint{0705.1158}.

\bibitem[{\citenamefont{Starobinsky}(2007)}]{Starobinsky:2007hu}
\bibinfo{author}{\bibfnamefont{A.~A.} \bibnamefont{Starobinsky}},
  \bibinfo{journal}{JETP Lett.} \textbf{\bibinfo{volume}{86}},
  \bibinfo{pages}{157} (\bibinfo{year}{2007}), \eprint{0706.2041}.

\bibitem[{\citenamefont{Scoccimarro}(2004)}]{Scoccimarro:2004tg}
\bibinfo{author}{\bibfnamefont{R.}~\bibnamefont{Scoccimarro}},
  \bibinfo{journal}{Phys. Rev.} \textbf{\bibinfo{volume}{D70}},
  \bibinfo{pages}{083007} (\bibinfo{year}{2004}), \eprint{astro-ph/0407214}.

\bibitem[{\citenamefont{Taruya et~al.}(2010)\citenamefont{Taruya, Nishimichi,
  and Saito}}]{Taruya:2010mx}
\bibinfo{author}{\bibfnamefont{A.}~\bibnamefont{Taruya}},
  \bibinfo{author}{\bibfnamefont{T.}~\bibnamefont{Nishimichi}},
  \bibnamefont{and} \bibinfo{author}{\bibfnamefont{S.}~\bibnamefont{Saito}},
  \bibinfo{journal}{Phys.Rev.} \textbf{\bibinfo{volume}{D82}},
  \bibinfo{pages}{063522} (\bibinfo{year}{2010}), \eprint{1006.0699}.

\bibitem[{\citenamefont{Nishimichi and Taruya}(2011)}]{Nishimichi:2011jm}
\bibinfo{author}{\bibfnamefont{T.}~\bibnamefont{Nishimichi}} \bibnamefont{and}
  \bibinfo{author}{\bibfnamefont{A.}~\bibnamefont{Taruya}},
  \bibinfo{journal}{Phys.Rev.} \textbf{\bibinfo{volume}{D84}},
  \bibinfo{pages}{043526} (\bibinfo{year}{2011}), \eprint{1106.4562}.

\bibitem[{\citenamefont{Matsubara}(2008)}]{Matsubara:2007wj}
\bibinfo{author}{\bibfnamefont{T.}~\bibnamefont{Matsubara}},
  \bibinfo{journal}{Phys. Rev.} \textbf{\bibinfo{volume}{D77}},
  \bibinfo{pages}{063530} (\bibinfo{year}{2008}), \eprint{0711.2521}.

\bibitem[{\citenamefont{Reid and White}(2009)}]{Reid:2011ar}
\bibinfo{author}{\bibfnamefont{B.~A.} \bibnamefont{Reid}} \bibnamefont{and}
  \bibinfo{author}{\bibfnamefont{M.}~\bibnamefont{White}},
  \bibinfo{journal}{Mon. Not. Roy. Astron. Soc.}
  \textbf{\bibinfo{volume}{417}}, \bibinfo{pages}{1913} (\bibinfo{year}{2009}),
  \eprint{1105.4165}.

\bibitem[{\citenamefont{{Carlson} et~al.}(2013)\citenamefont{{Carlson}, {Reid},
  and {White}}}]{Carlson:2012bu}
\bibinfo{author}{\bibfnamefont{J.}~\bibnamefont{{Carlson}}},
  \bibinfo{author}{\bibfnamefont{B.}~\bibnamefont{{Reid}}}, \bibnamefont{and}
  \bibinfo{author}{\bibfnamefont{M.}~\bibnamefont{{White}}},
  \bibinfo{journal}{Mon. Not. Roy. Astron. Soc.}
  \textbf{\bibinfo{volume}{429}}, \bibinfo{pages}{1674} (\bibinfo{year}{2013}),
  \eprint{1209.0780}.

\bibitem[{\citenamefont{Seljak and McDonald}(2011)}]{Seljak:2011tx}
\bibinfo{author}{\bibfnamefont{U.}~\bibnamefont{Seljak}} \bibnamefont{and}
  \bibinfo{author}{\bibfnamefont{P.}~\bibnamefont{McDonald}},
  \bibinfo{journal}{JCAP} \textbf{\bibinfo{volume}{1111}}, \bibinfo{pages}{039}
  (\bibinfo{year}{2011}), \eprint{1109.1888}.

\bibitem[{\citenamefont{Wang et~al.}(2013)\citenamefont{Wang, Reid, and
  White}}]{Wang:2013hwa}
\bibinfo{author}{\bibfnamefont{L.}~\bibnamefont{Wang}},
  \bibinfo{author}{\bibfnamefont{B.}~\bibnamefont{Reid}}, \bibnamefont{and}
  \bibinfo{author}{\bibfnamefont{M.}~\bibnamefont{White}},
  \bibinfo{journal}{Mon. Not. R. Astron. Soc.}  (\bibinfo{year}{2013}),
  \eprint{1306.1804}.

\bibitem[{\citenamefont{Vlah et~al.}(2012)\citenamefont{Vlah, Seljak, McDonald,
  Okumura, and Baldauf}}]{Vlah:2012ni}
\bibinfo{author}{\bibfnamefont{Z.}~\bibnamefont{Vlah}},
  \bibinfo{author}{\bibfnamefont{U.}~\bibnamefont{Seljak}},
  \bibinfo{author}{\bibfnamefont{P.}~\bibnamefont{McDonald}},
  \bibinfo{author}{\bibfnamefont{T.}~\bibnamefont{Okumura}}, \bibnamefont{and}
  \bibinfo{author}{\bibfnamefont{T.}~\bibnamefont{Baldauf}},
  \bibinfo{journal}{JCAP} \textbf{\bibinfo{volume}{1211}}, \bibinfo{pages}{009}
  (\bibinfo{year}{2012}), \eprint{1207.0839}.

\bibitem[{\citenamefont{Vlah et~al.}(2013)\citenamefont{Vlah, Seljak, Okumura,
  and Desjacques}}]{Vlah:2013lia}
\bibinfo{author}{\bibfnamefont{Z.}~\bibnamefont{Vlah}},
  \bibinfo{author}{\bibfnamefont{U.}~\bibnamefont{Seljak}},
  \bibinfo{author}{\bibfnamefont{T.}~\bibnamefont{Okumura}}, \bibnamefont{and}
  \bibinfo{author}{\bibfnamefont{V.}~\bibnamefont{Desjacques}}
  (\bibinfo{year}{2013}), \eprint{1308.6294}.

\bibitem[{\citenamefont{Reid et~al.}(2012)\citenamefont{Reid, Samushia, White,
  Percival, Manera et~al.}}]{Reid:2012sw}
\bibinfo{author}{\bibfnamefont{B.~A.} \bibnamefont{Reid}},
  \bibinfo{author}{\bibfnamefont{L.}~\bibnamefont{Samushia}},
  \bibinfo{author}{\bibfnamefont{M.}~\bibnamefont{White}},
  \bibinfo{author}{\bibfnamefont{W.~J.} \bibnamefont{Percival}},
  \bibinfo{author}{\bibfnamefont{M.}~\bibnamefont{Manera}},
  \bibnamefont{et~al.} (\bibinfo{year}{2012}), \eprint{1203.6641}.

\bibitem[{\citenamefont{Blake et~al.}(2011)\citenamefont{Blake, Davis, Poole,
  Parkinson, Brough et~al.}}]{Blake:2011wn}
\bibinfo{author}{\bibfnamefont{C.}~\bibnamefont{Blake}},
  \bibinfo{author}{\bibfnamefont{T.}~\bibnamefont{Davis}},
  \bibinfo{author}{\bibfnamefont{G.}~\bibnamefont{Poole}},
  \bibinfo{author}{\bibfnamefont{D.}~\bibnamefont{Parkinson}},
  \bibinfo{author}{\bibfnamefont{S.}~\bibnamefont{Brough}},
  \bibnamefont{et~al.}, \bibinfo{journal}{Mon.Not.Roy.Astron.Soc.}
  \textbf{\bibinfo{volume}{415}}, \bibinfo{pages}{2892} (\bibinfo{year}{2011}),
  \eprint{1105.2862}.

\bibitem[{\citenamefont{Fisher}(1995)}]{Fisher:1994ks}
\bibinfo{author}{\bibfnamefont{K.~B.} \bibnamefont{Fisher}},
  \bibinfo{journal}{Astrophys.J.} \textbf{\bibinfo{volume}{448}},
  \bibinfo{pages}{494} (\bibinfo{year}{1995}), \eprint{astro-ph/9412081}.

\bibitem[{\citenamefont{Hatton and Cole}(1998)}]{Hatton:1997xs}
\bibinfo{author}{\bibfnamefont{S.}~\bibnamefont{Hatton}} \bibnamefont{and}
  \bibinfo{author}{\bibfnamefont{S.}~\bibnamefont{Cole}},
  \bibinfo{journal}{Mon.Not.Roy.Astron.Soc.} \textbf{\bibinfo{volume}{296}},
  \bibinfo{pages}{10} (\bibinfo{year}{1998}), \eprint{astro-ph/9707186}.

\bibitem[{\citenamefont{Jennings et~al.}(2011)\citenamefont{Jennings, Baugh,
  and Pascoli}}]{Jennings:2010ne}
\bibinfo{author}{\bibfnamefont{E.}~\bibnamefont{Jennings}},
  \bibinfo{author}{\bibfnamefont{C.~M.} \bibnamefont{Baugh}}, \bibnamefont{and}
  \bibinfo{author}{\bibfnamefont{S.}~\bibnamefont{Pascoli}},
  \bibinfo{journal}{Astrophys.J.} \textbf{\bibinfo{volume}{727}},
  \bibinfo{pages}{L9} (\bibinfo{year}{2011}), \eprint{1011.2842}.

\bibitem[{\citenamefont{Taruya et~al.}(2013)\citenamefont{Taruya, Nishimichi,
  and Bernardeau}}]{Taruya:2013my}
\bibinfo{author}{\bibfnamefont{A.}~\bibnamefont{Taruya}},
  \bibinfo{author}{\bibfnamefont{T.}~\bibnamefont{Nishimichi}},
  \bibnamefont{and}
  \bibinfo{author}{\bibfnamefont{F.}~\bibnamefont{Bernardeau}}
  (\bibinfo{year}{2013}), \eprint{1301.3624}.

\bibitem[{\citenamefont{Ishikawa et~al.}(2013)\citenamefont{Ishikawa, Totani,
  Nishimichi, Takahashi, Yoshida et~al.}}]{Ishikawa:2013aea}
\bibinfo{author}{\bibfnamefont{T.}~\bibnamefont{Ishikawa}},
  \bibinfo{author}{\bibfnamefont{T.}~\bibnamefont{Totani}},
  \bibinfo{author}{\bibfnamefont{T.}~\bibnamefont{Nishimichi}},
  \bibinfo{author}{\bibfnamefont{R.}~\bibnamefont{Takahashi}},
  \bibinfo{author}{\bibfnamefont{N.}~\bibnamefont{Yoshida}},
  \bibnamefont{et~al.} (\bibinfo{year}{2013}), \eprint{1308.6087}.

\bibitem[{\citenamefont{Oka et~al.}(2013)\citenamefont{Oka, Saito, Nishimichi,
  Taruya, and Yamamoto}}]{Oka:2013cba}
\bibinfo{author}{\bibfnamefont{A.}~\bibnamefont{Oka}},
  \bibinfo{author}{\bibfnamefont{S.}~\bibnamefont{Saito}},
  \bibinfo{author}{\bibfnamefont{T.}~\bibnamefont{Nishimichi}},
  \bibinfo{author}{\bibfnamefont{A.}~\bibnamefont{Taruya}}, \bibnamefont{and}
  \bibinfo{author}{\bibfnamefont{K.}~\bibnamefont{Yamamoto}}
  (\bibinfo{year}{2013}), \eprint{1310.2820}.

\bibitem[{\citenamefont{Bernardeau et~al.}(2002)\citenamefont{Bernardeau,
  Colombi, Gaztanaga, and Scoccimarro}}]{Bernardeau:2001qr}
\bibinfo{author}{\bibfnamefont{F.}~\bibnamefont{Bernardeau}},
  \bibinfo{author}{\bibfnamefont{S.}~\bibnamefont{Colombi}},
  \bibinfo{author}{\bibfnamefont{E.}~\bibnamefont{Gaztanaga}},
  \bibnamefont{and}
  \bibinfo{author}{\bibfnamefont{R.}~\bibnamefont{Scoccimarro}},
  \bibinfo{journal}{Phys. Rept.} \textbf{\bibinfo{volume}{367}},
  \bibinfo{pages}{1} (\bibinfo{year}{2002}), \eprint{astro-ph/0112551}.

\bibitem[{\citenamefont{Kaiser}(1987)}]{Kaiser:1987qv}
\bibinfo{author}{\bibfnamefont{N.}~\bibnamefont{Kaiser}},
  \bibinfo{journal}{Mon. Not. Roy. Astron. Soc.}
  \textbf{\bibinfo{volume}{227}}, \bibinfo{pages}{1} (\bibinfo{year}{1987}).

\bibitem[{\citenamefont{{Hamilton}}(1992)}]{1992ApJ385L5H}
\bibinfo{author}{\bibfnamefont{A.~J.~S.} \bibnamefont{{Hamilton}}},
  \bibinfo{journal}{Astrophys. J.} \textbf{\bibinfo{volume}{385}},
  \bibinfo{pages}{L5} (\bibinfo{year}{1992}).

\bibitem[{\citenamefont{Taruya et~al.}(2009)\citenamefont{Taruya, Nishimichi,
  Saito, and Hiramatsu}}]{Taruya:2009ir}
\bibinfo{author}{\bibfnamefont{A.}~\bibnamefont{Taruya}},
  \bibinfo{author}{\bibfnamefont{T.}~\bibnamefont{Nishimichi}},
  \bibinfo{author}{\bibfnamefont{S.}~\bibnamefont{Saito}}, \bibnamefont{and}
  \bibinfo{author}{\bibfnamefont{T.}~\bibnamefont{Hiramatsu}},
  \bibinfo{journal}{Phys. Rev.} \textbf{\bibinfo{volume}{D80}},
  \bibinfo{pages}{123503} (\bibinfo{year}{2009}), \eprint{0906.0507}.

\bibitem[{\citenamefont{Okamura et~al.}(2011)\citenamefont{Okamura, Taruya, and
  Matsubara}}]{Okamura:2011nu}
\bibinfo{author}{\bibfnamefont{T.}~\bibnamefont{Okamura}},
  \bibinfo{author}{\bibfnamefont{A.}~\bibnamefont{Taruya}}, \bibnamefont{and}
  \bibinfo{author}{\bibfnamefont{T.}~\bibnamefont{Matsubara}},
  \bibinfo{journal}{JCAP} \textbf{\bibinfo{volume}{1108}}, \bibinfo{pages}{012}
  (\bibinfo{year}{2011}), \eprint{1105.1491}.

\bibitem[{\citenamefont{Crocce and Scoccimarro}(2008)}]{Crocce:2007dt}
\bibinfo{author}{\bibfnamefont{M.}~\bibnamefont{Crocce}} \bibnamefont{and}
  \bibinfo{author}{\bibfnamefont{R.}~\bibnamefont{Scoccimarro}},
  \bibinfo{journal}{Phys. Rev.} \textbf{\bibinfo{volume}{D77}},
  \bibinfo{pages}{023533} (\bibinfo{year}{2008}), \eprint{0704.2783}.

\bibitem[{\citenamefont{Crocce et~al.}(2012)\citenamefont{Crocce, Scoccimarro,
  and Bernardeau}}]{Crocce:2012fa}
\bibinfo{author}{\bibfnamefont{M.}~\bibnamefont{Crocce}},
  \bibinfo{author}{\bibfnamefont{R.}~\bibnamefont{Scoccimarro}},
  \bibnamefont{and}
  \bibinfo{author}{\bibfnamefont{F.}~\bibnamefont{Bernardeau}},
  \bibinfo{journal}{Mon. Not. Roy. Astron. Soc.}
  \textbf{\bibinfo{volume}{427}}, \bibinfo{pages}{2537} (\bibinfo{year}{2012}),
  \eprint{1207.1465}.

\bibitem[{\citenamefont{Taruya et~al.}(2012)\citenamefont{Taruya, Bernardeau,
  Nishimichi, and Codis}}]{Taruya:2012ut}
\bibinfo{author}{\bibfnamefont{A.}~\bibnamefont{Taruya}},
  \bibinfo{author}{\bibfnamefont{F.}~\bibnamefont{Bernardeau}},
  \bibinfo{author}{\bibfnamefont{T.}~\bibnamefont{Nishimichi}},
  \bibnamefont{and} \bibinfo{author}{\bibfnamefont{S.}~\bibnamefont{Codis}},
  \bibinfo{journal}{Phys.Rev.} \textbf{\bibinfo{volume}{D86}},
  \bibinfo{pages}{103528} (\bibinfo{year}{2012}), \eprint{1208.1191}.

\bibitem[{\citenamefont{Crocce and Scoccimarro}(2006)}]{Crocce:2005xy}
\bibinfo{author}{\bibfnamefont{M.}~\bibnamefont{Crocce}} \bibnamefont{and}
  \bibinfo{author}{\bibfnamefont{R.}~\bibnamefont{Scoccimarro}},
  \bibinfo{journal}{Phys. Rev.} \textbf{\bibinfo{volume}{D73}},
  \bibinfo{pages}{063519} (\bibinfo{year}{2006}), \eprint{astro-ph/0509418}.

\bibitem[{\citenamefont{Carlson et~al.}(2009)\citenamefont{Carlson, White, and
  Padmanabhan}}]{Carlson:2009it}
\bibinfo{author}{\bibfnamefont{J.}~\bibnamefont{Carlson}},
  \bibinfo{author}{\bibfnamefont{M.}~\bibnamefont{White}}, \bibnamefont{and}
  \bibinfo{author}{\bibfnamefont{N.}~\bibnamefont{Padmanabhan}},
  \bibinfo{journal}{Phys. Rev.} \textbf{\bibinfo{volume}{D80}},
  \bibinfo{pages}{043531} (\bibinfo{year}{2009}), \eprint{0905.0479}.

\bibitem[{\citenamefont{Bernardeau et~al.}(2012)\citenamefont{Bernardeau,
  Taruya, and Nishimichi}}]{Bernardeau:2012ux}
\bibinfo{author}{\bibfnamefont{F.}~\bibnamefont{Bernardeau}},
  \bibinfo{author}{\bibfnamefont{A.}~\bibnamefont{Taruya}}, \bibnamefont{and}
  \bibinfo{author}{\bibfnamefont{T.}~\bibnamefont{Nishimichi}}
  (\bibinfo{year}{2012}), \eprint{1211.1571}.

\bibitem[{\citenamefont{Saito et~al.}(2011)\citenamefont{Saito, Takada, and
  Taruya}}]{Saito:2010pw}
\bibinfo{author}{\bibfnamefont{S.}~\bibnamefont{Saito}},
  \bibinfo{author}{\bibfnamefont{M.}~\bibnamefont{Takada}}, \bibnamefont{and}
  \bibinfo{author}{\bibfnamefont{A.}~\bibnamefont{Taruya}},
  \bibinfo{journal}{Phys.Rev.} \textbf{\bibinfo{volume}{D83}},
  \bibinfo{pages}{043529} (\bibinfo{year}{2011}), \eprint{1006.4845}.

\bibitem[{\citenamefont{{Zhao} et~al.}(2013)\citenamefont{{Zhao}, {Saito},
  {Percival}, {Ross}, {Montesano}, {Viel}, {Schneider}, {Manera},
  {Miralda-Escud{\'e}}, {Palanque-Delabrouille} et~al.}}]{Zhao:2012xw}
\bibinfo{author}{\bibfnamefont{G.-B.} \bibnamefont{{Zhao}}},
  \bibinfo{author}{\bibfnamefont{S.}~\bibnamefont{{Saito}}},
  \bibinfo{author}{\bibfnamefont{W.~J.} \bibnamefont{{Percival}}},
  \bibinfo{author}{\bibfnamefont{A.~J.} \bibnamefont{{Ross}}},
  \bibinfo{author}{\bibfnamefont{F.}~\bibnamefont{{Montesano}}},
  \bibinfo{author}{\bibfnamefont{M.}~\bibnamefont{{Viel}}},
  \bibinfo{author}{\bibfnamefont{D.~P.} \bibnamefont{{Schneider}}},
  \bibinfo{author}{\bibfnamefont{M.}~\bibnamefont{{Manera}}},
  \bibinfo{author}{\bibfnamefont{J.}~\bibnamefont{{Miralda-Escud{\'e}}}},
  \bibinfo{author}{\bibfnamefont{N.}~\bibnamefont{{Palanque-Delabrouille}}},
  \bibnamefont{et~al.}, \bibinfo{journal}{Mon. Not. Roy. Astron. Soc.}
  \textbf{\bibinfo{volume}{436}}, \bibinfo{pages}{2038} (\bibinfo{year}{2013}),
  \eprint{1211.3741}.

\bibitem[{\citenamefont{Koyama et~al.}(2009)\citenamefont{Koyama, Taruya, and
  Hiramatsu}}]{Koyama:2009me}
\bibinfo{author}{\bibfnamefont{K.}~\bibnamefont{Koyama}},
  \bibinfo{author}{\bibfnamefont{A.}~\bibnamefont{Taruya}}, \bibnamefont{and}
  \bibinfo{author}{\bibfnamefont{T.}~\bibnamefont{Hiramatsu}}
  (\bibinfo{year}{2009}), \eprint{0902.0618}.

\bibitem[{\citenamefont{Dvali et~al.}(2000)\citenamefont{Dvali, Gabadadze, and
  Porrati}}]{Dvali:2000hr}
\bibinfo{author}{\bibfnamefont{G.~R.} \bibnamefont{Dvali}},
  \bibinfo{author}{\bibfnamefont{G.}~\bibnamefont{Gabadadze}},
  \bibnamefont{and} \bibinfo{author}{\bibfnamefont{M.}~\bibnamefont{Porrati}},
  \bibinfo{journal}{Phys. Lett.} \textbf{\bibinfo{volume}{B485}},
  \bibinfo{pages}{208} (\bibinfo{year}{2000}), \eprint{hep-th/0005016}.

\bibitem[{\citenamefont{Hiramatsu and Taruya}(2009)}]{Hiramatsu:2009ki}
\bibinfo{author}{\bibfnamefont{T.}~\bibnamefont{Hiramatsu}} \bibnamefont{and}
  \bibinfo{author}{\bibfnamefont{A.}~\bibnamefont{Taruya}}
  (\bibinfo{year}{2009}), \eprint{0902.3772}.

\bibitem[{\citenamefont{Bernardeau et~al.}(2008)\citenamefont{Bernardeau,
  Crocce, and Scoccimarro}}]{Bernardeau:2008fa}
\bibinfo{author}{\bibfnamefont{F.}~\bibnamefont{Bernardeau}},
  \bibinfo{author}{\bibfnamefont{M.}~\bibnamefont{Crocce}}, \bibnamefont{and}
  \bibinfo{author}{\bibfnamefont{R.}~\bibnamefont{Scoccimarro}},
  \bibinfo{journal}{Phys. Rev.} \textbf{\bibinfo{volume}{D78}},
  \bibinfo{pages}{103521} (\bibinfo{year}{2008}), \eprint{0806.2334}.

\bibitem[{\citenamefont{Marchini and Salvatelli}(2013)}]{Marchini:2013oya}
\bibinfo{author}{\bibfnamefont{A.}~\bibnamefont{Marchini}} \bibnamefont{and}
  \bibinfo{author}{\bibfnamefont{V.}~\bibnamefont{Salvatelli}},
  \bibinfo{journal}{Phys.Rev.} \textbf{\bibinfo{volume}{D88}},
  \bibinfo{pages}{027502} (\bibinfo{year}{2013}), \eprint{1307.2002}.

\bibitem[{\citenamefont{Lombriser et~al.}(2012)\citenamefont{Lombriser, Slosar,
  Seljak, and Hu}}]{Lombriser:2010mp}
\bibinfo{author}{\bibfnamefont{L.}~\bibnamefont{Lombriser}},
  \bibinfo{author}{\bibfnamefont{A.}~\bibnamefont{Slosar}},
  \bibinfo{author}{\bibfnamefont{U.}~\bibnamefont{Seljak}}, \bibnamefont{and}
  \bibinfo{author}{\bibfnamefont{W.}~\bibnamefont{Hu}},
  \bibinfo{journal}{Phys.Rev.} \textbf{\bibinfo{volume}{D85}},
  \bibinfo{pages}{124038} (\bibinfo{year}{2012}), \eprint{1003.3009}.

\bibitem[{\citenamefont{Yamamoto et~al.}(2010)\citenamefont{Yamamoto, Nakamura,
  Hutsi, Narikawa, and Sato}}]{Yamamoto:2010ie}
\bibinfo{author}{\bibfnamefont{K.}~\bibnamefont{Yamamoto}},
  \bibinfo{author}{\bibfnamefont{G.}~\bibnamefont{Nakamura}},
  \bibinfo{author}{\bibfnamefont{G.}~\bibnamefont{Hutsi}},
  \bibinfo{author}{\bibfnamefont{T.}~\bibnamefont{Narikawa}}, \bibnamefont{and}
  \bibinfo{author}{\bibfnamefont{T.}~\bibnamefont{Sato}},
  \bibinfo{journal}{Phys.Rev.} \textbf{\bibinfo{volume}{D81}},
  \bibinfo{pages}{103517} (\bibinfo{year}{2010}), \eprint{1004.3231}.

\bibitem[{\citenamefont{Okada et~al.}(2013)\citenamefont{Okada, Totani, and
  Tsujikawa}}]{Okada:2012mn}
\bibinfo{author}{\bibfnamefont{H.}~\bibnamefont{Okada}},
  \bibinfo{author}{\bibfnamefont{T.}~\bibnamefont{Totani}}, \bibnamefont{and}
  \bibinfo{author}{\bibfnamefont{S.}~\bibnamefont{Tsujikawa}},
  \bibinfo{journal}{Phys.Rev.} \textbf{\bibinfo{volume}{D87}},
  \bibinfo{pages}{103002} (\bibinfo{year}{2013}), \eprint{1208.4681}.

\bibitem[{\citenamefont{Schmidt
  et~al.}(2009{\natexlab{a}})\citenamefont{Schmidt, Vikhlinin, and
  Hu}}]{Schmidt:2009am}
\bibinfo{author}{\bibfnamefont{F.}~\bibnamefont{Schmidt}},
  \bibinfo{author}{\bibfnamefont{A.}~\bibnamefont{Vikhlinin}},
  \bibnamefont{and} \bibinfo{author}{\bibfnamefont{W.}~\bibnamefont{Hu}},
  \bibinfo{journal}{Phys.Rev.} \textbf{\bibinfo{volume}{D80}},
  \bibinfo{pages}{083505} (\bibinfo{year}{2009}{\natexlab{a}}),
  \eprint{0908.2457}.

\bibitem[{\citenamefont{Li et~al.}(2013)\citenamefont{Li, Hellwing, Koyama,
  Zhao, Jennings et~al.}}]{Li:2012by}
\bibinfo{author}{\bibfnamefont{B.}~\bibnamefont{Li}},
  \bibinfo{author}{\bibfnamefont{W.~A.} \bibnamefont{Hellwing}},
  \bibinfo{author}{\bibfnamefont{K.}~\bibnamefont{Koyama}},
  \bibinfo{author}{\bibfnamefont{G.-B.} \bibnamefont{Zhao}},
  \bibinfo{author}{\bibfnamefont{E.}~\bibnamefont{Jennings}},
  \bibnamefont{et~al.}, \bibinfo{journal}{Mon.Not.Roy.Astron.Soc.}
  \textbf{\bibinfo{volume}{428}}, \bibinfo{pages}{743} (\bibinfo{year}{2013}),
  \eprint{1206.4317}.

\bibitem[{\citenamefont{Jennings et~al.}(2012)\citenamefont{Jennings, Baugh,
  Li, Zhao, and Koyama}}]{Jennings:2012pt}
\bibinfo{author}{\bibfnamefont{E.}~\bibnamefont{Jennings}},
  \bibinfo{author}{\bibfnamefont{C.~M.} \bibnamefont{Baugh}},
  \bibinfo{author}{\bibfnamefont{B.}~\bibnamefont{Li}},
  \bibinfo{author}{\bibfnamefont{G.-B.} \bibnamefont{Zhao}}, \bibnamefont{and}
  \bibinfo{author}{\bibfnamefont{K.}~\bibnamefont{Koyama}},
  \bibinfo{journal}{Mon.Not.Roy.Astron.Soc.} \textbf{\bibinfo{volume}{425}},
  \bibinfo{pages}{2128} (\bibinfo{year}{2012}), \eprint{1205.2698}.

\bibitem[{\citenamefont{Li et~al.}(2012{\natexlab{a}})\citenamefont{Li, Zhao,
  Teyssier, and Koyama}}]{Li:2011vk}
\bibinfo{author}{\bibfnamefont{B.}~\bibnamefont{Li}},
  \bibinfo{author}{\bibfnamefont{G.-B.} \bibnamefont{Zhao}},
  \bibinfo{author}{\bibfnamefont{R.}~\bibnamefont{Teyssier}}, \bibnamefont{and}
  \bibinfo{author}{\bibfnamefont{K.}~\bibnamefont{Koyama}},
  \bibinfo{journal}{JCAP} \textbf{\bibinfo{volume}{1201}}, \bibinfo{pages}{051}
  (\bibinfo{year}{2012}{\natexlab{a}}), \eprint{1110.1379}.

\bibitem[{\citenamefont{Teyssier}(2002)}]{Teyssier:2001cp}
\bibinfo{author}{\bibfnamefont{R.}~\bibnamefont{Teyssier}},
  \bibinfo{journal}{Astron.Astrophys.} \textbf{\bibinfo{volume}{385}},
  \bibinfo{pages}{337} (\bibinfo{year}{2002}), \eprint{astro-ph/0111367}.

\bibitem[{\citenamefont{Nishimichi et~al.}(2009)}]{Nishimichi:2008ry}
\bibinfo{author}{\bibfnamefont{T.}~\bibnamefont{Nishimichi}}
  \bibnamefont{et~al.}, \bibinfo{journal}{Publ. Astron. Soc. Jap.}
  \textbf{\bibinfo{volume}{61}}, \bibinfo{pages}{321} (\bibinfo{year}{2009}),
  \eprint{0810.0813}.

\bibitem[{\citenamefont{Taruya et~al.}(2011)\citenamefont{Taruya, Saito, and
  Nishimichi}}]{Taruya:2011tz}
\bibinfo{author}{\bibfnamefont{A.}~\bibnamefont{Taruya}},
  \bibinfo{author}{\bibfnamefont{S.}~\bibnamefont{Saito}}, \bibnamefont{and}
  \bibinfo{author}{\bibfnamefont{T.}~\bibnamefont{Nishimichi}},
  \bibinfo{journal}{Phys.Rev.} \textbf{\bibinfo{volume}{D83}},
  \bibinfo{pages}{103527} (\bibinfo{year}{2011}), \eprint{1101.4723}.

\bibitem[{\citenamefont{Lam et~al.}(2012)\citenamefont{Lam, Nishimichi,
  Schmidt, and Takada}}]{Lam:2012by}
\bibinfo{author}{\bibfnamefont{T.~Y.} \bibnamefont{Lam}},
  \bibinfo{author}{\bibfnamefont{T.}~\bibnamefont{Nishimichi}},
  \bibinfo{author}{\bibfnamefont{F.}~\bibnamefont{Schmidt}}, \bibnamefont{and}
  \bibinfo{author}{\bibfnamefont{M.}~\bibnamefont{Takada}},
  \bibinfo{journal}{Phys.Rev.Lett.} \textbf{\bibinfo{volume}{109}},
  \bibinfo{pages}{051301} (\bibinfo{year}{2012}), \eprint{1202.4501}.

\bibitem[{\citenamefont{Lam et~al.}(2013)\citenamefont{Lam, Schmidt,
  Nishimichi, and Takada}}]{Lam:2013kma}
\bibinfo{author}{\bibfnamefont{T.~Y.} \bibnamefont{Lam}},
  \bibinfo{author}{\bibfnamefont{F.}~\bibnamefont{Schmidt}},
  \bibinfo{author}{\bibfnamefont{T.}~\bibnamefont{Nishimichi}},
  \bibnamefont{and} \bibinfo{author}{\bibfnamefont{M.}~\bibnamefont{Takada}},
  \bibinfo{journal}{Phys.Rev.} \textbf{\bibinfo{volume}{D88}},
  \bibinfo{pages}{023012} (\bibinfo{year}{2013}), \eprint{1305.5548}.

\bibitem[{\citenamefont{Takahashi et~al.}(2009)}]{Takahashi:2009bq}
\bibinfo{author}{\bibfnamefont{R.}~\bibnamefont{Takahashi}}
  \bibnamefont{et~al.}, \bibinfo{journal}{Astrophys. J.}
  \textbf{\bibinfo{volume}{700}}, \bibinfo{pages}{479} (\bibinfo{year}{2009}),
  \eprint{0902.0371}.

\bibitem[{\citenamefont{Takahashi et~al.}(2011)\citenamefont{Takahashi,
  Yoshida, Takada, Matsubara, Sugiyama et~al.}}]{Takahashi:2009ty}
\bibinfo{author}{\bibfnamefont{R.}~\bibnamefont{Takahashi}},
  \bibinfo{author}{\bibfnamefont{N.}~\bibnamefont{Yoshida}},
  \bibinfo{author}{\bibfnamefont{M.}~\bibnamefont{Takada}},
  \bibinfo{author}{\bibfnamefont{T.}~\bibnamefont{Matsubara}},
  \bibinfo{author}{\bibfnamefont{N.}~\bibnamefont{Sugiyama}},
  \bibnamefont{et~al.}, \bibinfo{journal}{Astrophys.J.}
  \textbf{\bibinfo{volume}{726}}, \bibinfo{pages}{7} (\bibinfo{year}{2011}),
  \eprint{0912.1381}.

\bibitem[{\citenamefont{Cole et~al.}(2005)}]{Cole:2005sx}
\bibinfo{author}{\bibfnamefont{S.}~\bibnamefont{Cole}} \bibnamefont{et~al.}
  (\bibinfo{collaboration}{The 2dFGRS}), \bibinfo{journal}{Mon. Not. Roy.
  Astron. Soc.} \textbf{\bibinfo{volume}{362}}, \bibinfo{pages}{505}
  (\bibinfo{year}{2005}), \eprint{astro-ph/0501174}.

\bibitem[{\citenamefont{Sanchez and Cole}(2008)}]{Sanchez:2007rc}
\bibinfo{author}{\bibfnamefont{A.~G.} \bibnamefont{Sanchez}} \bibnamefont{and}
  \bibinfo{author}{\bibfnamefont{S.}~\bibnamefont{Cole}},
  \bibinfo{journal}{Mon.Not.Roy.Astron.Soc.} \textbf{\bibinfo{volume}{385}},
  \bibinfo{pages}{830} (\bibinfo{year}{2008}), \eprint{0708.1517}.

\bibitem[{\citenamefont{Brax and Valageas}(2012)}]{Brax:2012sy}
\bibinfo{author}{\bibfnamefont{P.}~\bibnamefont{Brax}} \bibnamefont{and}
  \bibinfo{author}{\bibfnamefont{P.}~\bibnamefont{Valageas}},
  \bibinfo{journal}{Phys.Rev.} \textbf{\bibinfo{volume}{D86}},
  \bibinfo{pages}{063512} (\bibinfo{year}{2012}), \eprint{1205.6583}.

\bibitem[{\citenamefont{Brax and Valageas}(2013)}]{Brax:2013fna}
\bibinfo{author}{\bibfnamefont{P.}~\bibnamefont{Brax}} \bibnamefont{and}
  \bibinfo{author}{\bibfnamefont{P.}~\bibnamefont{Valageas}}
  (\bibinfo{year}{2013}), \eprint{1305.5647}.

\bibitem[{\citenamefont{Schmidt
  et~al.}(2009{\natexlab{b}})\citenamefont{Schmidt, Lima, Oyaizu, and
  Hu}}]{Schmidt:2008tn}
\bibinfo{author}{\bibfnamefont{F.}~\bibnamefont{Schmidt}},
  \bibinfo{author}{\bibfnamefont{M.~V.} \bibnamefont{Lima}},
  \bibinfo{author}{\bibfnamefont{H.}~\bibnamefont{Oyaizu}}, \bibnamefont{and}
  \bibinfo{author}{\bibfnamefont{W.}~\bibnamefont{Hu}},
  \bibinfo{journal}{Phys.Rev.} \textbf{\bibinfo{volume}{D79}},
  \bibinfo{pages}{083518} (\bibinfo{year}{2009}{\natexlab{b}}),
  \eprint{0812.0545}.

\bibitem[{\citenamefont{Lombriser et~al.}(2013)\citenamefont{Lombriser, Li,
  Koyama, and Zhao}}]{Lombriser:2013wta}
\bibinfo{author}{\bibfnamefont{L.}~\bibnamefont{Lombriser}},
  \bibinfo{author}{\bibfnamefont{B.}~\bibnamefont{Li}},
  \bibinfo{author}{\bibfnamefont{K.}~\bibnamefont{Koyama}}, \bibnamefont{and}
  \bibinfo{author}{\bibfnamefont{G.-B.} \bibnamefont{Zhao}},
  \bibinfo{journal}{Phys.Rev.} \textbf{\bibinfo{volume}{D87}},
  \bibinfo{pages}{123511} (\bibinfo{year}{2013}), \eprint{1304.6395}.

\bibitem[{\citenamefont{Ferraro et~al.}(2011)\citenamefont{Ferraro, Schmidt,
  and Hu}}]{Ferraro:2010gh}
\bibinfo{author}{\bibfnamefont{S.}~\bibnamefont{Ferraro}},
  \bibinfo{author}{\bibfnamefont{F.}~\bibnamefont{Schmidt}}, \bibnamefont{and}
  \bibinfo{author}{\bibfnamefont{W.}~\bibnamefont{Hu}},
  \bibinfo{journal}{Phys.Rev.} \textbf{\bibinfo{volume}{D83}},
  \bibinfo{pages}{063503} (\bibinfo{year}{2011}), \eprint{1011.0992}.

\bibitem[{\citenamefont{Li et~al.}(2012{\natexlab{b}})\citenamefont{Li, Zhao,
  and Koyama}}]{Li:2011pj}
\bibinfo{author}{\bibfnamefont{B.}~\bibnamefont{Li}},
  \bibinfo{author}{\bibfnamefont{G.-B.} \bibnamefont{Zhao}}, \bibnamefont{and}
  \bibinfo{author}{\bibfnamefont{K.}~\bibnamefont{Koyama}},
  \bibinfo{journal}{Mon.Not.Roy.Astron.Soc.} \textbf{\bibinfo{volume}{421}},
  \bibinfo{pages}{3481} (\bibinfo{year}{2012}{\natexlab{b}}),
  \eprint{1111.2602}.

\bibitem[{\citenamefont{{Fontanot} et~al.}(2013)\citenamefont{{Fontanot},
  {Puchwein}, {Springel}, and {Bianchi}}}]{Bianchi:2013foa}
\bibinfo{author}{\bibfnamefont{F.}~\bibnamefont{{Fontanot}}},
  \bibinfo{author}{\bibfnamefont{E.}~\bibnamefont{{Puchwein}}},
  \bibinfo{author}{\bibfnamefont{V.}~\bibnamefont{{Springel}}},
  \bibnamefont{and}
  \bibinfo{author}{\bibfnamefont{D.}~\bibnamefont{{Bianchi}}},
  \bibinfo{journal}{Mon. Not. Roy. Astron. Soc.}
  \textbf{\bibinfo{volume}{436}}, \bibinfo{pages}{2672} (\bibinfo{year}{2013}),
  \eprint{1307.5065}.

\end{thebibliography}
\bibliographystyle{apsrev}


\end{document}